\documentclass[aps,preprint,groupedaddress]{revtex4}
\usepackage{graphicx}

\begin{document}

\title{Magnetization orientation dependence of the quasiparticle
spectrum and hysteresis in ferromagnetic metal nanoparticles}


\author{A.~Cehovin} 
\affiliation{Division of Solid State Theory, Department of Physics,
Lund University, SE-223 62 Lund, Sweden}
\author{C.M.~Canali} 
\affiliation{Department of Technology, Kalmar University, 391 82 Kalmar, 
Sweden,
and Division of Solid State Theory, Department of Physics, 
Lund University, SE-223 62 Lund, Sweden}
\author{A.H.~MacDonald}
\affiliation{Department of Physics, University of Texas at Austin, 
Austin TX 78712}


\date{\today}

\begin{abstract}

We use a microscopic Slater-Koster tight-binding model with short-range
exchange and atomic spin-orbit interactions that realistically captures generic 
features of ferromagnetic metal nanoparticles to address the mesoscopic physics of magnetocrystalline
anisotropy and hysteresis in nanoparticle quasiparticle
excitation spectra.  
Our analysis is based on qualitative arguments supported by 
self-consistent Hartree-Fock calculations for 
nanoparticles containing up to 260 
atoms.  Calculations of the total energy as a function 
of magnetization direction 
demonstrate that the magnetic anisotropy per atom fluctuates by  
several percents when the number of electrons in the particle changes by one, even for 
the largest particles we consider.  Contributions of individual orbitals to
the magnetic anisotropy are characterized by a broad distribution with a mean
more than two orders of magnitude smaller than its variance and with no
detectable correlations between anisotropy contribution and quasiparticle 
energy.  We find that the discrete quasiparticle excitation spectrum 
of a nanoparticle displays a complex non-monotonic dependence
on an external magnetic field, with abrupt jumps
when the magnetization direction is reversed by
the field, explaining recent spectroscopic studies of magnetic nanoparticles.
Our results suggests the existence of a broad cross-over from a weak spin-orbit 
coupling to a strong spin-orbit coupling regime, occurring over the range from
approximately 200- to 1000-atom nanoparticles.

\end{abstract}

\maketitle

\section{Introduction}
Interest in the properties of magnetic
nanoparticles has grown recently, partly because of advances in synthesis and 
measurement techniques
and partly because of potential applications for high storage-density magnetic media
and spin electronics.  Ferromagnetic nanoparticles with diameters of a
few nanometers containing of order
of 1000 or fewer atoms can now be reliably fabricated and studied 
with a variety of different methods\cite{gueron1999, black2000, jamet2001}.
Small monodomain magnetic particles have 
traditionally\cite{stoner_wohlfart, zijlstra} 
been described using classical {\em micromagnetic} theory, in which
the total energy is expressed as a function
of the magnetization orientation.  Shape and 
magnetocrystalline magnetic anisotropy leads to a dependence of energy on orientation, 
to barriers that separate minima that occur at {\em easy} magnetization
orientations\cite{skomski}, and to hysteretic discontinuous 
changes in orientation as a function
of the strength of an external magnetic field.  When the size of a 
magnetic particle is only a few nanometers, the discrete nature of its quantum
energy spectrum can be directly observable at low temperatures and
starts to affect the magnetic properties of the particle.

A milestone in the experimental study of {\it individual}
ferromagnetic metal nanoparticles was achieved recently 
by Deshmukh, Gu\'eron, Ralph {\it et.al.}\cite{gueron1999,deshmukh2001}, 
using single-electron tunneling spectroscopy.
By exploiting the Coulomb blockade effect, these experimentalists
were able to resolve {\it individual quantum states} in the discrete many-body excitation 
spectrum of single ferromagnetic metal nanograins with sizes in the range from $1$ to $4$ nm. 
As in bulk ferromagnetic metals, the low-energy excitations of the nanoparticles that 
were probed in these experiments involve both particle-hole excitations of the
electronic quasiparticles and quantized collective excitations of the 
magnetization-orientation field that appears in the classical 
micromagnetic theory.  In an initial attempt to achieve an understanding of 
the novel physics evident in the 
external field dependence of the experimental excitation spectra, 
two of us recently\cite{cmc_ahm2000prl} analyzed a simple quantum 
model with long range exchange 
interactions.  
We were able to demonstrate explicitly that the low-energy excitations of a 
ferromagnetic metal nanoparticle are specified by the occupation numbers of its quasiparticle orbitals,
as in a Fermi liquid, {\em and} by the global orientation of the aligned spins
of all single-occupied (and therefore spin-polarized) orbitals. 
This model is not, however, able to account realistically for the
influence of spin-orbit interactions, which play the essential role 
in controlling the complex hysteretic 
behavior seen in these experiments\cite{gueron1999,deshmukh2001}.
Kleff, von Delft, Deshmukh and Ralph\cite{kleff2001prl,kleff_vdelft2001prb} 
have proposed that the single-electron tunneling spectroscopy experiments
can be explained by accounting for non-equilibrium spin accumulation, 
and by assuming that the magnetic anisotropy energy of a small magnetic 
particle has surprisingly large fluctuations as a function of the number of 
electrons on the particle.  This assumption leads to 
a non-trivial magnetic field dependence of tunneling resonances 
that resembles experimental behavior. More critically, these authors point out
that if non-equilibrium spin and quasi-particle
excitations both occur\cite{kleff_vdelft2001prb},
the low energy spectra are characterized by many closely spaced 
resonances, consistent with experiment\cite{kleff_vdelft2001prb}.
In Ref.~\onlinecite{ahm_cmc2001ssc} we presented a possible approach 
toward achieving a unified and consistent quantum description 
of both collective and quasiparticle
physics in magnetic nanoparticles.  

The attempts put forward in 
Refs.~\onlinecite{cmc_ahm2000prl,kleff2001prl,kleff_vdelft2001prb} 
to develop a quantum description of ferromagnetic metal nanoparticles do not
address the microscopic origin of the
magnetic anisotropy and the fluctuations of this quantity that are a 
necessary consequence of spin-orbit interactions in mesoscopic systems. 
This article addresses these issues and, more in general, investigates
the changes in 
magnetic properties of small metallic nanoparticles that occur because of the 
finite spacing between quasiparticle levels near the Fermi energy.  
Our conclusions follow 
from qualitative arguments based on perturbation theory expressions
for the influence of spin-orbit interactions on quasiparticle energy levels,
and on numerical studies of a simplified model that 
we believe is sufficiently realistic to describe 
generic aspects of the interrelated mesoscopic physics 
of quasiparticle energy-levels
and magnetic anisotropy energy in ferromagnetic metal nanoparticles.
We are particularly interested in the variation of quasiparticle 
energies with fields 
on the scale of the coercive field, which is trivial in the absence of 
spin-orbit interactions, but entirely non-trivial in their presence.  
The model we study is based on a Slater-Koster tight binding Hamiltonian and a 
mean-field treatment of exchange interactions.  
Our aim is to understand the dependence of magnetic anisotropy and  
quasiparticle energy-level-spacing statistics on 
particle size and shape, external magnetic field, and,
with single-electron-transistor systems in mind, also on electron number.
We find that,
because of the absence of strong correlations between 
angular momentum operator matrix elements and orbital energy differences,
the size of spin-orbit induced energy shifts in nanoparticles and 
bulk perfect crystals 
are similar.  On the basis of estimates for energy shifts and 
for avoided crossing gaps,
we predict that a crossover from weak to strong quasiparticle 
spin-orbit scattering
will occur over the range from approximately 200-atom to 
approximately 1000-atom nanoparticles. 
We find that for small particles 
the contribution of individual quasiparticle orbitals to the magnetocrystalline 
anisotropy energy has a wide distribution, characterized by a variance 
comparable to the spin-orbit-scattering lifetime broadening energy, 
$\hbar \tau_{SO}^{-1}$,
and a mean that is smaller by more than two orders of magnitude. 
Surprisingly, we find no measurable correlation 
between the contributions to magnetic anisotropy
from quasiparticles that are close in energy in this limit.   
As a result of the statistical properties of the quasiparticle
magnetic anisotropies, the total magnetic anisotropy per atom fluctuates
by several percents when the number of electrons in the nanoparticle
changes by one, even for particles containing 260 atoms.
Finally, in agreement with experiment\cite{gueron1999,deshmukh2001}, 
we find that the quasiparticle excitation spectrum 
exhibits a complex non-monotonic behavior as a function of an
external magnetic field, with abrupt jumps
when the magnetization orientation of a nanoparticle 
changes discontinuously in response to the field. 
Our analysis provides insight into mesoscopic fluctuations of 
the micromagnetic energy functional which appears in the classical
theory of a magnetic nanoparticle.

Our paper is organized in the following way.
In Section~\ref{model} we introduce the model and describe the formalism.
In Section~\ref{so_anis} we analyze the qualitative 
change in quasiparticle energy-level
statistics induced by spin-orbit interactions, 
and discuss the connection between the quasiparticle properties and 
the magnetic anisotropy of a ferromagnetic nanoparticle.
Fluctuations of the magnetic anisotropy as a function of spin-splitting
field, atom number, and electron number are investigated in 
Section~\ref{fluct}.  Magnetic hysteresis and the 
external field dependence of quasi-particle energy levels are discussed in 
Section~\ref{hysteresis}. Finally, in Section~\ref{conclusion} 
we summarize our
findings and present our conclusions. 

\section{The Model}
\label{model}

We model the nano-particle as a cluster of ${\cal N}_a$ atoms located 
on the sites of a truncated crystal.  The numerical results we present here
are for a cobalt cluster whose truncated f.c.c. crystal is 
circumscribed by a hemisphere whose equator lies in the $xy$-plane of the 
f.c.c. crystal \footnote{\label{fnote1}
The choice of a hemisphere is motivated by the tunneling experiments
of Ref.~\cite{gueron1999,deshmukh2001}. 
The crystal structure could not be identified
in these experiments. High resolution transmission electron microscopy
measurements on Co clusters prepared in a different way\cite{jamet2001} 
have shown that the nanoparticles are well crystallized in the f.c.c. structure
even though bulk Co has an h.c.p. structure.}. 

\subsection{Tight-binding Hamiltonian and Slater-Koster parameters}
The model we use is intended to qualitatively capture the 
physics of a transition-metal, itinerant-electron ferromagnet.
We use a s-p-d tight-binding model for the quasiparticle orbitals,
with 18 orbitals per atom including the spin-degree of 
freedom.  Nine orbitals per Co atom are occupied for in neutral nanoparticles.  
The full Hamiltonian is, 

\begin{equation}
{\cal H}={\cal H}_{\rm band}+{\cal H}_{\rm exch}+{\cal H}_{\rm SO}+
{\cal H}_{\rm Zee}\; .
\label{totham}
\end{equation}

Here ${\cal H}_{\rm band}$ is a one-body term describing the orbital
motion of the electrons. In second quantization, it has the form
\begin{equation}
{\cal H}_{\rm band} = \sum_{i,j} \sum_{s} \sum_{\mu_1, \mu_2}
t^{i,j}_{\mu_1,\mu_2,s}
c^{\dagger}_{i,\mu_1,s}c^{\phantom{\dagger}}_{j,\mu_2,s}
\label{H1b}
\end{equation}
where $c^{\dagger}$ and $c^{\phantom{\dagger}}$ are Fermion creation
and annihilation operators for single-particle states labeled by
$i,\mu, s$.
The indices
$i,j$ are atomic site labels, and $t^{i,j}$ couples up to  
second nearest-neighbors.
The indices
$\mu_1,\mu_2$ label the nine distinct atomic orbitals 
(one $4s$, three $4p$ and
five $3d$). The spin degrees of
freedom, labeled by the index $s$, double the number of
orbitals at each site.  It is useful for us to vary the spin-quantization axis,
which is specified by a  unit vector $\hat \Omega(\Theta, \Phi)$ where $\Theta$ and 
$\Phi$ are the usual angular coordinates defined with respect to the f.c.c. crystal
axes.  The parameters $t^{i,j}_{\mu_1,\mu_2,s}$ are
Slater-Koster parameters\cite{slater_koster} obtained after performing a
L\"owdin symmetric orthogonalization procedure\cite{mattheiss}
on the set of Slater-Koster parameters for non-orthogonal atomic orbitals 
of bulk spin-unpolarized Co\cite{papaconstantopoulos}. 

The electron-electron interaction term in Eq.~(\ref{totham}) is simplified 
by introducing explicitly
\footnote{\label{fnote2}Notice however, that the one
body term, $H_{\rm band}$ implicitly includes a mean-field approximation
to those portions of the interaction not captured by the exchange term.}
only the ferromagnetic exchange interaction, ${\cal H}_{\rm exch}$, between
the electrons spins of $d$-orbitals on the same atomic site.   These interactions 
are largely responsible for magnetic order in transition metal ferromagnets:  

\begin{equation}
{\cal H}_{\rm exch}=
-2U_{dd}\sum_{i} \vec{S}_{d,i} \cdot \vec{S}_{d,i}\; ,
\label{exchH}
\end{equation}
where
\begin{equation}
\vec{S}_{d,i} = \sum_{\mu\in d } \vec S_{i,\mu}
= \sum_{\mu\in d } 
\frac{1}{2}
\sum_{s,s'}c^{\dagger}_{i,\mu,s}{\vec\tau}^{\phantom{\dagger}}_{s,s'}
c^{\phantom{\dagger}}_{i,\mu,s'}\, .
\label{pauli}
\end{equation}
The parameter $U_{dd}$ in Eq.~(\ref{exchH}) determines 
the strength of the exchange interaction
and is set equal to $1$ eV. 

This value of $U_{dd}$ gives rise in our finite clusters
to an average magnetic moment per atom 
of the order of $2 \mu_{\rm B}$, which is larger than
the bulk value in Co\cite{papaconstantopoulos}, 
in agreement with other calculations\cite{duan2001} 
and experimental results\cite{billas1994, billas1997} for Co clusters.
This value of $U_{dd}$ is also approximately consistent
with a mean-field 
relationship, derived below, between band spin-splitting and magnetization,
using bulk values\cite{papaconstantopoulos} for these two quantities.
In Eq.~(\ref{pauli}), $\vec\tau$ is a vector whose components
$\tau^{\alpha}$, $\alpha=x,y,z$ are the three Pauli matrices.

The third term in Eq.~(\ref{totham}), ${\cal H}_{\rm SO}$, 
is a one-body operator, essentially atomic in character, 
representing the spin-orbit interaction. It can be written as\cite{bruno89}
\begin{equation}
{\cal H}_{\rm SO} = \xi_d\sum_i \sum_{\mu,\mu',s,s'}
\langle \mu,s| {\vec L}\cdot {\vec S}|\mu',s'\rangle
c^{\dagger}_{i,\mu,s}c^{\phantom{\dagger}}_{i,\mu',s'}
\end{equation}

The atomic matrix elements
$\langle \mu,s| {\vec L}\cdot {\vec S}|\mu',s'\rangle \equiv
\langle i, \mu,s| {\vec L}\cdot {\vec S}|i, \mu',s'\rangle$
can be been calculated explicitly
as a function of the direction of the magnetization 
$\hat \Omega$\cite{SOmatrix_elements}.
The energy scale $\xi_d$, which characterizes the coupling between
spin and orbital degrees of freedom,  varies in the range from 50
to 100 meV in bulk 3d transition metal ferromagnets\cite{soinmetals}.
For our calculations we have used $\xi_d =82$ meV, taken from 
Ref.~\cite{pastor1998}. 

Spin-orbit coupling gives rise to a dependence of the total
energy of a ferromagnet on the direction of its spontaneous magnetization, an effect known
as magneto-crystalline anisotropy\footnote{\label{fnote3}
More precisely, this
is the magneto-electric contribution to the magneto-crystalline anisotropy.};
with spin-orbit coupling the magnetization is partially orbital
in character\cite{ericsson1990, jansen1995} 
and is sensitive to orbital anisotropy due to 
crystal-field interactions on an atomic site, due to the spatial arrangement of 
the atoms neighboring that site to which electrons can hop, and due to the overall shape
of the full nanoparticle that becomes available to an electron after many hops
\footnote{\label{fnote4}
The additional anisotropy due to magnetic dipole interactions is also sensitive
to the overall nanoparticle shape, but is not included here because it is
generally less important for nanoparticles than for macroscopic samples.  
This contribution
to the anisotropy is additive to a good approximation, does not have substantial
mesoscopic fluctuations, and can be simply added to the effects 
discussed here when
it is important.}.   We will see below that the hemispherical shape of the 
nanoparticles we have studied plays the dominant role in determining 
their magnetic anisotropy energy. 

The last term in Eq.~(\ref{totham}) is a local one-body operator,
representing the Zeeman coupling
of the orbital and spin degrees of freedom to an external 
magnetic field $\vec{H}_{\rm ext}$:

\begin{eqnarray}
{\cal H}_{\rm Zee}= &&-\mu_{\rm B}\sum_i \sum_{\mu,\mu',s,s'}
\langle \mu,s|(\vec{L}+g_{s}\vec{S}|\mu',s')\rangle
\cdot\vec{H}_{\rm ext}\, 
c^{\dagger}_{i,\mu,s}c^{\phantom{\dagger}}_{i,\mu',s'}\nonumber\\
 =&& -\mu_{\rm B}\sum_i \vec{H}_{\rm ext}\cdot \Big\{\sum_{\mu,\mu',s}
\langle \mu,s|\vec{L}|\mu',s)\rangle
\, c^{\dagger}_{i,\mu,s}c^{\phantom{\dagger}}_{i,\mu',s}
+ {g_s\over 2}
\sum_{\mu,s,s'}c^{\dagger}_{i,\mu,s}{\vec\tau}^{\phantom{\dagger}}_{s,s'}
c^{\phantom{\dagger}}_{i,\mu,s'}\Big\}\; .
\end{eqnarray}

The extreme sensitivity of magnetization orientation to external magnetic field
is a combined effect of the collective behavior of many electrons in a ferromagnetic
nanoparticle and the smallness of the magnetic anisotropy energy relative to
the overall energy gain associated with ferromagnetic order.  The most delicate physics 
of a ferromagnetic nanoparticle, and in our view the most interesting, is that associated with 
magnetization direction reorientation by weak external magnetic fields.   Thus the 
interplay between the Zeeman term and the magnetic anisotropy produced by spin-orbit interactions
is at the heart of the physics we intend to address. 

\subsection{Mean-field approximation}
\label{hartree_fock}

We seek a ferromagnetic solution to the mean-field equations for this model,
decoupling the quartic term in the exchange interaction using the ansatz 
\begin{equation}
\vec{S}_{d,i} = \langle \vec{S}_{d,i}\rangle + \delta \vec{S}_{d,i}\; , 
\label{decoup}
\end{equation}
ignoring terms that are second order in $\delta \vec{S}_{d,i}\;$, 
and determining ground state expectation values, $\langle \dots \rangle$, self-consistently.
This standard procedure leads to a Hamiltonian that can be diagonalized numerically
and to a self-consistency condition that can be solved iteratively to 
determine the mean-field order parameters.
For the present calculation we have simplified this procedure further,
by averaging the spin-splitting exchange mean field, 
\begin{equation}
\vec h_{i} \equiv h \hat \Omega  = 
\frac{U_{dd}}{g_s \mu_B} 2 \langle \vec{S}_{d,i}\rangle 
\end{equation}
over all sites.  Our motivation for doing so is to simplify the magnetic anisotropy
energy landscape discussed below, forcing all spins to change their orientations coherently.  
We recognize that complicated non-collinear 
spin configurations\cite{car1998, aleks_SC2001} 
commonly occur in magnetic nanoparticles,
and that under the action of an external field small groups of atoms
can change their orientation relative to other parts of the nanoparticle. 
Complex magnetization reorientation processes are obvious in addition
spectroscopy experiments\cite{deshmukh2001}.  By forcing atoms to change 
their magnetic orientations coherently, we are restricting our 
attention to relatively large nanoparticles (${\cal N}_a > 50$) in which most atoms 
have parallel spins and to nanoparticles with simple bistable hysteretic behavior 
in which the physics we address will be easier to study.  
The beautiful series of 
detailed SQUID magnetometry experiments by 
Wernsdorfer and colleagues\cite{wernsdorfer97, jamet2001} 
demonstrate that nanoparticles can be prepared that do have simple coherent 
magnetization reversal properties. 

By using a simplified model Hamiltonian we are able to 
deal with larger nanoparticle systems than would be possible with 
a first principles calculation\cite{pastor1995, pastor1998, car1998,duan2001};
since we are interested only in generic aspects of the ferromagnetic 
nanoparticle physics there is little to gain from the additional realism that 
could be achieved by performing 
self-consistent spin-density functional calculations. 

Given the averaged spin-splitting field $\vec h$, the mean-field
Hamiltonian is now a single-body operator 

\begin{equation}
{\cal H}_{\rm MF}(\vec h)
={\cal H_{\rm band}}+ {\cal H_{\rm SO}}+{\cal H_{\rm Zee}}+ 
{\vec h\cdot \vec h \over 2U_{dd}} (g_s\mu_{\rm B})^2 N_{a}
- g_s\mu_{\rm B}\vec{h}\cdot \sum_{i}2\vec{S}_{d,i}\; . 
\end{equation}
The self-consistent spin-spitting field is also the field,
denoted by $\vec h^{\star}$ below, at which   
the total ground-state energy function 
$E(\vec h) = \langle {\cal H}_{\rm MF}(\vec h)\rangle$ is minimized.
Notice that $g_s\mu_{\rm B}h^{\star}$ has the dimension of an energy. In fact,
in absence of spin-orbit interaction, $2g_s\mu_{\rm B}h^{\star}$ 
can be identified
with the band spin-splitting
$\Delta = \epsilon_{{\rm F}a}^0 - \epsilon_{{\rm F}i}^0$,
where $\epsilon_{{\rm F}a}^0$ and $\epsilon_{{\rm F}i}^0$ are
the majority and minority spin quasi-particle Fermi energies.
\footnote{\label{fnote5}
The use of two Fermi energies for majority and minority spins
can generate some confusion: in this terminology,
the energies are pure kinetic (or band) energies,
{\it i.e } the exchange contribution is not included. When the exchange
energy is included, as we do below in the context of mean-field theory,
majority and minority spins have the same Fermi level.}
Diagonalization of the quadratic Hamiltonian 
yields a set of quasi-particle energies $\{\epsilon_{n,s}\},\; n= 1,2, \dots$, 
which in the ground state are filled up to a Fermi energy determined by the number of
valence electrons in the nanoparticle. Strictly speaking, because of the spin-orbit
interaction, the spin character of the corresponding eigenstates 
$|\psi_{n,s}\rangle $ is not well defined.

For small nanoparticles, most eigenstates have {\it predominantly} spin-up or spin-down character
and we sometimes use this property to assign spin labels $s = 
\; \uparrow$ for spins along the order direction and $s = \; \downarrow$ for reversed 
spins.

\section{Spin-orbit interactions and the magnetic anisotropy energy}
\label{so_anis}

The magnetic anisotropy energy
of small ferromagnetic particles\cite{jamet2001} 
has two fundamentally distinct origins, long-range magnetic-dipole interactions which 
cause a dependence on overall sample shape and short-range exchange interactions
that, because of atomic-like spin-orbit interactions, 
are sensitive to all aspects of the electron hopping network including 
bulk crystal symmetry, facet orientations, and also overall sample shape.
We concentrate here on spin-orbit-induced magnetocrystalline anisotropy 
which gives rise to the most interesting physics in ferromagnetic nanoparticles.
When magnetostatic shape anisotropy is important, it can be added as a separate 
contribution.  We begin our discussion with some qualitative estimates of the 
effect of spin-orbit interactions that are based on 
perturbation theory\cite{bruno88, bruno90}. 

\subsection{Perturbation theory considerations}
\label{perturb}

In bulk 3d transition metal ferromagnets, spin-orbit
interactions are relatively weak. Their coupling strength  
is less than $10 \%$ of the d-band width $W_d$ in bulk materials\cite{soinmetals},
allowing the energy shifts they produce to be estimated  
perturbatively.  Because of angular momentum quenching in the absence of external fields, 
the expectation value of
$H_{\rm SO}$ is zero, even in case of ferromagnets\cite{soinmetals}.

The quasi-particle energy shift due to spin-orbit interactions
is given by second-order perturbation theory as 
\begin{equation}
\epsilon_{\rm SO}\equiv \epsilon_{n,s} - \epsilon^0_{n,s} 
= {(\xi_d)^2\over4}
\sum_{s'\atop\scriptstyle m\neq n} 
{|\langle \psi^0_{m,s'}|\vec L|\psi^0_{n,s}\rangle
\cdot \vec \tau_{s',s}|^2\over 
\epsilon^0_{n,s}-\epsilon^0_{m,s'}}\; ,
\label{eq:perttheory} 
\end{equation}
where $|\psi^0_{n,s}\rangle$ and $\epsilon^0_{n,s}$
are respectively the single-particle eigenstates and energies
in the absence of spin-orbit interaction. 
In small particles,
the importance of the spin-orbit interactions can be assessed by
comparing the spin-orbit energy shift $\epsilon_{\rm SO}$ with the single-particle
mean-level spacing $\delta$. 

In an infinite periodic solid only states at the same $\vec k$ are 
coupled and these are separated energetically by an energy comparable
to the bandwidth $W_d$.
In a nanoparticle a given state will be coupled to many 
other orbitals, but the coupling matrix elements will be reduced in
accord with the following sum rule: 

\begin{equation}
\sum_{s'\atop\scriptstyle m\neq n} |\langle \psi^0_{m,s'}|\vec L|\psi^0_{n,s}\rangle \cdot 
\vec \tau_{s',s} |^2 
= \langle \psi^0_{n,s} | \vec L \cdot \vec L | \psi^0_{n,s} \rangle  \sim 4. 
\label{eq:sumrule} 
\end{equation} 

The estimate for the right hand side of Eq.~\ref{eq:sumrule} is based on the
atomic character of the angular momentum in our model, and uses that
$\vec L \cdot \vec L \sim  [5*6+ 3*2+ 1*0]/9$,
with the estimate
of the typical $\vec L \cdot \vec L $ expectation value representing an average
over $d$, $s$ and $p$ orbitals.
It follows that, unless there are important correlations between 
angular momentum matrix elements and quasiparticle energy differences, 
the typical shift in energy caused by the spin-orbit interaction is 
$ \epsilon _{\rm SO} \sim {(\xi_d)^2\over W_d}$, in both bulk crystals and 
in nanoparticles,

which is in the range between 1 to 10 meV.
For example in Co, using $\xi_d = 82$meV and $W_d \sim 5 $eV, this 
rough estimate gives for the magnitude of the spin-orbit energy shift
$ \epsilon _{\rm SO}$ $\sim 1.3$meV.  The sign of the shift might be expected
to be sensitive to the spacing of nearby quasiparticle orbitals. 
The anisotropy energy, that is, the dependence of the total band energy
on the magnetization orientation, is given to a good approximation by a 
partial canceling sum of spin-orbit induced energy shift dependences on magnetic orientation.

In the approximation that the exchange field is orbital independent, majority
spin and minority spin orbitals are identical and differ only in their occupation
numbers.  In this approximation there is no contribution to the anisotropy energy from doubly
occupied orbitals. 
Because of the cancellations, the anisotropy energy per atom is much
smaller than  $\epsilon _{\rm SO}$.
For example, the zero-temperature anisotropy energy per atom 
in bulk is $60 \mu$eV for h.c.p. Co,
and $\approx 1 \mu$eV for b.c.c. Fe and f.c.c. Ni\cite{stearns1986}. 
In a finite or disordered system, there will
always be perturbative coupling to quasiparticle states close in energy 
in Eq.~\ref{eq:perttheory},
but the matrix elements, which satisfy the sum rule of Eq.~\ref{eq:sumrule},
will be distributed among many states and typical energy shifts in nanoparticles should generally 
be comparable to those in bulk perfect crystals.  
Typically net anisotropy energies per atom
in small magnetic particles are larger than the bulk because of the loss 
of symmetry at the surface. 

An important quantity used to characterize the strength of 
spin-orbit interactions in bulk systems and large nanoparticles is the 
the spin-orbit scattering time, 
$\tau_{\rm SO}$\cite{halperin1986, brouwer2000, matveev2000, petta2001}.

In the weak coupling regime, it is given by the Fermi's golden rule
\begin{equation}
\hbar\tau_{\rm SO}^{-1}= 
{(\xi_d)^2\over4}
\sum_{s'\atop\scriptstyle m\neq n}
|\langle \psi^0_{m,s'}|\vec L|\psi^0_{n,s}\rangle
\cdot \vec \tau_{s',s}|^2
\delta[\epsilon^0_{n,s}-\epsilon^0_{m,s'}]\; ,
\label{eq:so_sc_time}
\end{equation}

where the $\delta-$function is understood to be broadened to a 
width much larger than the level spacing.  Assuming that there is no
correlation between angular momentum matrix elements and orbital 
energy differences, it follows from the sum rule mentioned above
that 
\begin{equation}
\hbar \tau_{\rm SO}^{-1} \sim \epsilon _{\rm SO}\ \sim {\xi_d^2\over W_d}\;.
\end{equation}

The absence of strong correlations between energy differences and angular momentum
matrix elements, a property that we find somewhat surprising, has
been verified numerically as we discuss below.  The character of the 
nanoparticle quasiparticle energy spectrum changes when these intensive energy 
scales become comparable to the nanoparticle level spacing $\delta$.   

\subsection{Numerical Results for a Co nanoparticle}

 \begin{figure}
 \includegraphics[width=11.cm,height=11.cm]{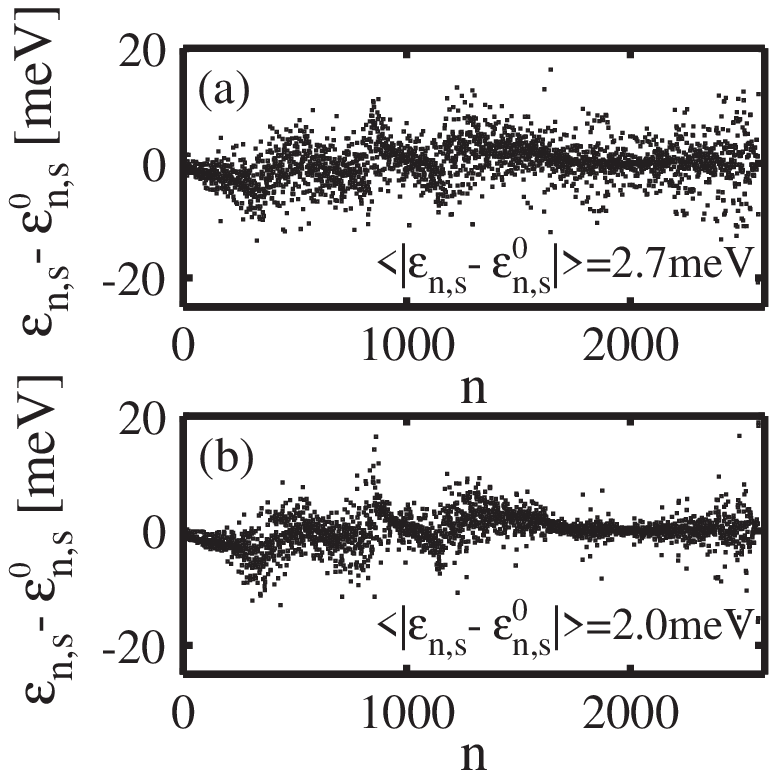}%
 \caption{Single-particle energy shifts caused by spin-orbit interactions
in a ferromagnetic Cobalt nanoparticle of 143 atoms. The single-particle
mean level spacing
at the Fermi energy is $\delta \approx 4.3$ meV. 
(a) Magnetization
in the z-direction. (b) Magnetization in the x-direction.}
 \label{soeig.eps}
 \end{figure}

The qualitative considerations of the previous section
provide a framework for thinking about the effects of spin-orbit interactions.
Our microscopic model, on the other hand, allows us to explore realistic 
magnetic nanoparticle systems in great detail.
We have studied numerically nanoparticles containing up to 260 atoms.
Most of the results presented below are for hemispherical 
143-atom nanoparticles. 
The calculations have been performed with a 
R12000 300MHz processor on an SGI Origin 2000 
computer. Diagonalizations rely on LAPACK 
drivers. A single diagonalization 
of the Hamiltonian for a 143-atom cluster has a running time of
approximately 1 hour, and requires around 750Mb 
of internal memory.
In Fig.~[\ref{soeig.eps}] we plot the energy shifts
caused by spin-orbit interaction for 
a hemispherical Cobalt nanoparticle of 
143 atoms with a f.c.c. crystal structure.
For this nanoparticle size, the minority and majority single-particle mean-level spacings
at the Fermi level are $\delta_{\downarrow}\approx 4.9$ meV and
$\delta_{\uparrow}\approx 50$ meV respectively, when spin-orbit interactions
are absent.  The single-particle mean-level spacing averaged over all states 
(i.e. without distinguishing between majority and minority levels) 
is $\delta \approx 4.3$ meV at the Fermi level\footnote{\label{fnote5b}
This is approximately the value 
of the single-particle mean-level at the Fermi level
also when spin-orbit interaction is included.}.
The spin-orbit induced shifts are both positive 
and negative and their absolute values go
from 1 meV up to 10 meV. The average absolute value of the energy 
shifts is 2.7 meV when the magnetization is in the z-direction,
and 2 meV when the magnetization is in the $xy$-plane, 
consistent with the rough estimates
above.  We also note that there is not a strong correlation between the sign
of the energy shift and the energy of the orbital.

 \begin{figure}
 \includegraphics[width=12.cm,height=6.cm]{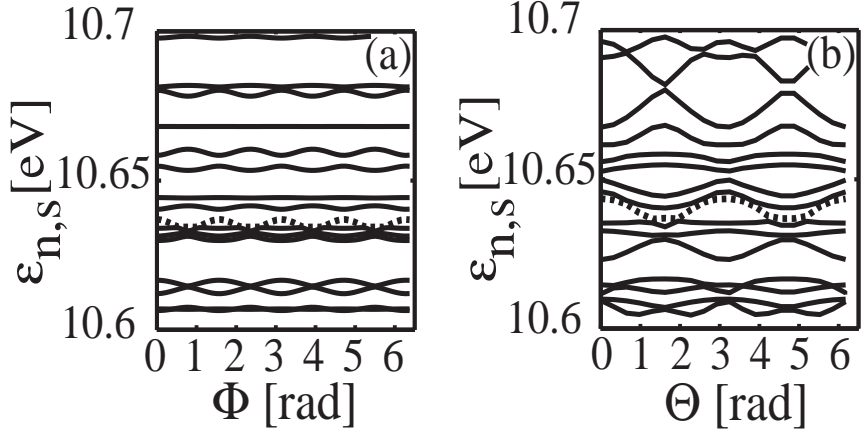}%
 \caption{Variation of a few quasiparticle energies 
of a 143-atom Co nanoparticle 
as a function of the direction of the magnetization $\hat \Omega$.
The Fermi level is the dotted line.
(a) $\hat \Omega$ lies in the $xy$-plane and $\Phi$ is the angle with $x$-axis. 
(b) $\hat \Omega$ lies in the $zx$-plane and $\Theta$ is the angle with the $z$-axis.
}
 \label{aniseigcomb}
 \end{figure}

Because of the spin-orbit interaction, each individual eigenlevel 
has an energy dependence on the 
spin-splitting field (or magnetization) direction $\hat \Omega$. 
To illustrate typical properties of these dependences, we plot 
in Fig.~[\ref{aniseigcomb}] the variation of
a few energy levels around the Fermi level for cases in which
the magnetization rotates in $zx$-plane and $xy$-planes respectively.
In the absence of spin-orbit interactions there would be no dependence of
any of these orbital energies on magnetization orientation.

 \begin{figure}
 \includegraphics[width=11.cm,height=11.cm]{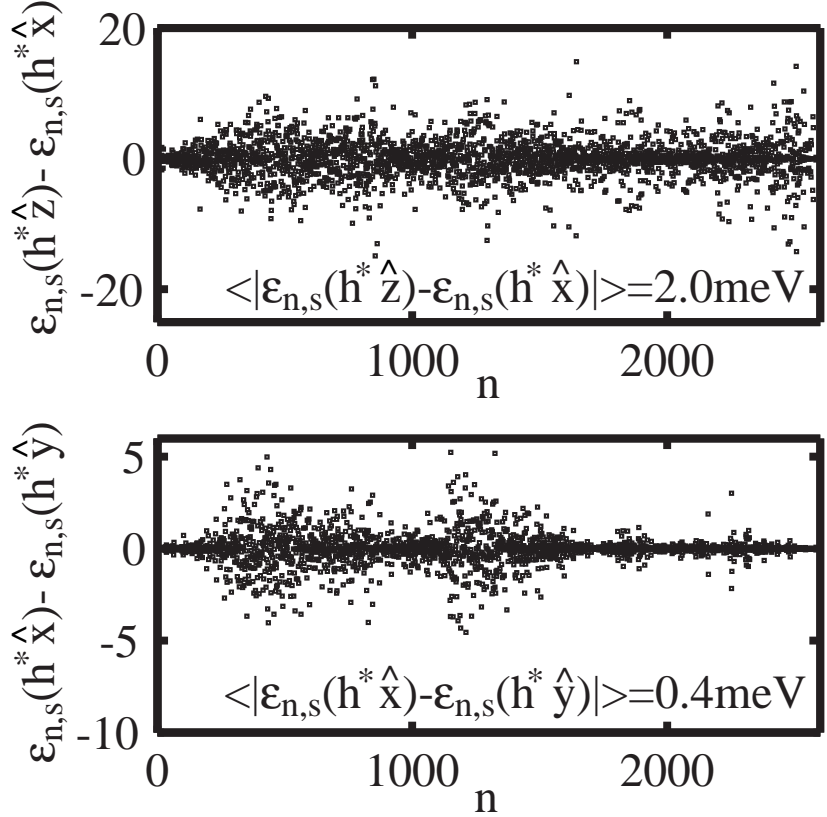}%
 \caption{Single-particle level anisotropies in the $xy$- and $zx$-planes
for a 143-atom nanoparticle.}
 \label{eiganis3}
 \end{figure}

\begin{figure}
 \includegraphics[width=11.cm,height=11.cm]{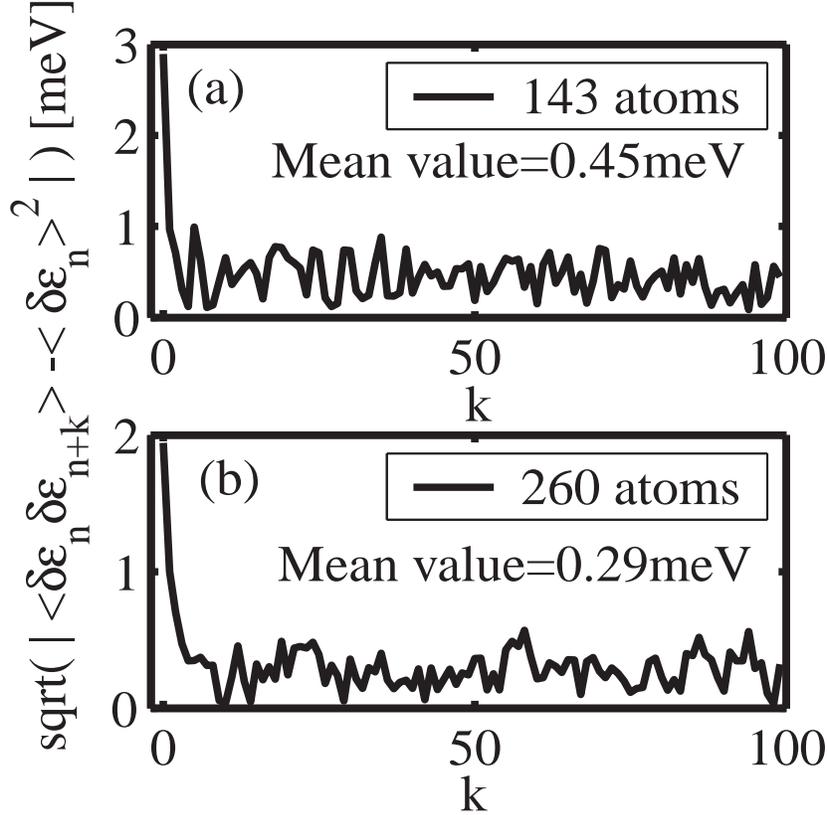}%
 \caption{Correlation function of single-particle anisotropies as defined
in Eq.~[\ref{corr_f}]. The correlation function drops immediately to zero
for $k>0$.
Note that the $k=0$ value of the correlation function is equal to
the width of the
anisotropy distribution, plotted in Fig.~[\ref{histeiganis2}].}
 \label{correlation}
 \end{figure}

Notice that the angle dependence in the $xy$-plane is considerably weaker
than in the $zx$-plane, for the hemispherical nanoparticle we consider. 
This trend indicates that for our nanoparticles, it is the overall
shape which dominates the spin-orbit induced anisotropy
physics.  For this size cluster, there are many narrowly avoided level 
crossings, a point to which we return below.
The difference in eigenvalue sensitivity to magnetization rotations
in the two planes is clearly visible in
Fig.~[\ref{eiganis3}], where we plot the single-particle level anisotropies
$\epsilon_{n,s}(h^{\star}_{\hat z}\,\hat z)- \epsilon_{n,s}
(h^{\star}_{\hat x}\,\hat x)$
and $\epsilon_{n,s}(h^{\star}_{\hat x}\,\hat x)- 
\epsilon_{n,s}(h^{\star}_{\hat y}\,\hat y)$
versus the eigenvalue index $n$.
Here $h^{\star}_{\hat x}$ and $h^{\star}_{\hat z}$ 
are the magnitudes of the self-consistent
spin-splitting field when its direction is along $\hat x$ and $\hat z$
respectively.  
We note that typical change in orbital energy between $\hat x$ and 
$\hat z$ direction magnetizations is only $\sim 30\%$ smaller than the 
typical total shifts induced by spin-orbit interactions.  On the other hand
the typical difference in orbital energy between $\hat x$ 
and $\hat y$ direction 
magnetization is five times smaller than the corresponding spin-orbit induced 
energy shift.  
All orbital energies are relatively insensitive to the magnetization
orientation within the $xy$-plane.  
In both cases the correlation between
position within the band and the sign and magnitude of the energy shift is weak.
In addition, energy shifts at nearby energies are weakly correlated. 
That is, the correlation function
\begin{equation}
\langle \delta \epsilon_{n,s} \delta \epsilon_{n+k,s'}\rangle -
\langle \delta \epsilon_{n,s}\rangle^2\; ,
\ \ \ \ \delta \epsilon_{n,s} = \epsilon_{n,s}(h_1^{\star}\hat\Omega_1)-
\epsilon_{n,s}(h_2^{\star}\hat\Omega_2)\; ,
\label{corr_f}
\end{equation}
where the average $\langle ....\rangle$
is over the occupied levels $n$, drops to zero very rapidly with $k$,
as clearly shown in Fig.~[\ref{correlation}].

 \begin{figure}
 \includegraphics[width=11.cm,height=11.cm]{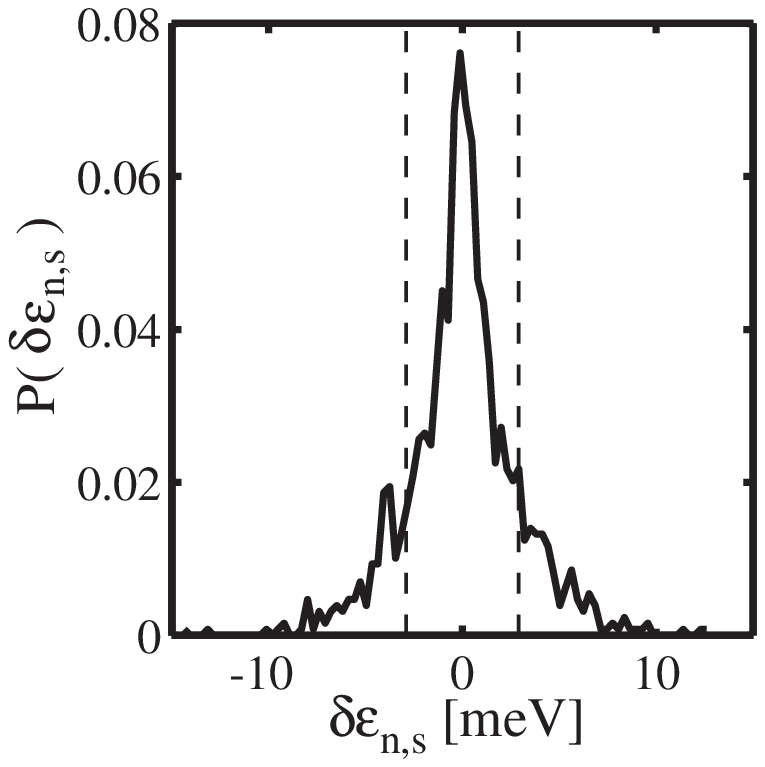}%
 \caption{Distribution function of single-particle anisotropies
in the $zx$-plane, $\delta\epsilon_{n,s}=\big[\epsilon_{n,s}(h^{\star}\hat z) -
\epsilon_{n,s}(h^{\star}\hat x)\big]$, for a 143-atom nanoparticle.
The mean value is $\langle\delta\epsilon_{n,s}\rangle=15 \mu$eV, 
the width of the
distribution (enclosed by the vertical dashed lines)
is $\Delta(\delta\epsilon_{n,s}) =2.9$ meV.}

 \label{histeiganis2}
 \end{figure}

It is useful to consider the {\it distribution} of the quasiparticle
anisotropies, $P(\delta\epsilon_{n,s})$.
As an example, in Fig.~\ref{histeiganis2} we plot the distribution of
anisotropies in the
$zx$-plane, $\delta\epsilon_{n,s}=\big[\epsilon_{n,s}(h^{\star}\hat z) -
\epsilon_{n,s}(h^{\star}\hat x)\big]$, 
constructed
with the $N= 9\times{\cal N}_a=9\times143$ 
occupied single-particle states of a 143-atom nanoparticle.
The distribution has a width -- characterized by the root mean square --
$\Delta(\delta\epsilon_{n,s}) \sim 2.2\,\xi_d^2 / W_d \sim 2.9$ meV 
and a much smaller mean value
$ \langle \delta\epsilon_{n,s}\rangle =
\langle \epsilon_{n,s}(h^{\star}\hat z) -
\epsilon_{n,s}(h^{\star}\hat x)\rangle \sim 15 \mu$eV.
Note that $\Delta(\delta\epsilon_{n,s})$ is exactly equal to the $k=0$ value
of the correlation function displayed in Fig.~[\ref{correlation}].
Single particle anisotropies of groups of orbitals over within a
specified energy range tend to be anti-correlated, leading to typical
averages smaller than the variance of the distribution.
The large difference between the distribution mean and variance will play
an important role in Sec.~\ref{anis_fluctsN}, where we discuss fluctuations 
of the anisotropy energy as a function of electron number.

The width of the distribution, $\Delta(\delta\epsilon_{n,s})$,
gives a measure of the average magnitude of 
the single-particle level anisotropy, and the ratio 
$\Delta(\delta\epsilon_{n,s})/\delta$
characterizes the strength of mixing between quasiparticle orbitals that 
results from spin-orbit interactions.
As mentioned previously, this identification of weak and strong spin-orbit interaction 
regimes is equivalent to 
the usual one\cite{halperin1986, brouwer2000, matveev2000, petta2001} 
based on the comparison of the spin-orbit scattering time $\tau_{\rm SO}$
and the mean-level spacing $\delta$.

When $\delta \tau_{\rm SO}/\hbar >> 1$,
a limit achieved for small enough particle size for any value of $\xi_d$,  
spin-orbit coupling is a relatively weak effect and there is little
mixing between spin-up and spin-down states. As a consequence,
the level crossings between states of predominantly opposite
spins that occur as a function of the magnitude and orientation 
of $\vec h$, will be only weakly avoided.
With increasing particle size, $\delta$ decreases and we enter
the regime of strong-spin orbit interaction and strong level
repulsion. The single-particle spectrum becomes relatively rigid,
level crossing will be strongly avoided, and $\delta$ 
will limit the variation of individual
levels as a function of the magnetization direction.

Within our model we have found that the cross-over between these
two regimes is very broad, it starts for nanoparticles containing of order
200 atoms and, as we argue below, it will be completed when the nanoparticles
contain approximately 1000 atom.
For a 143-atom nanoparticle, we are already approaching
the cross-over regime; we find  
$\Delta(\delta\epsilon_{n,s})\sim 2.9 $meV
while the single-particle mean-level spacing 
at the Fermi level is $\delta \approx 4.3$ meV.
In this regime level crossings will 
be moderately avoided.

 \begin{figure}
 \includegraphics[width=11.cm,height=11.cm]{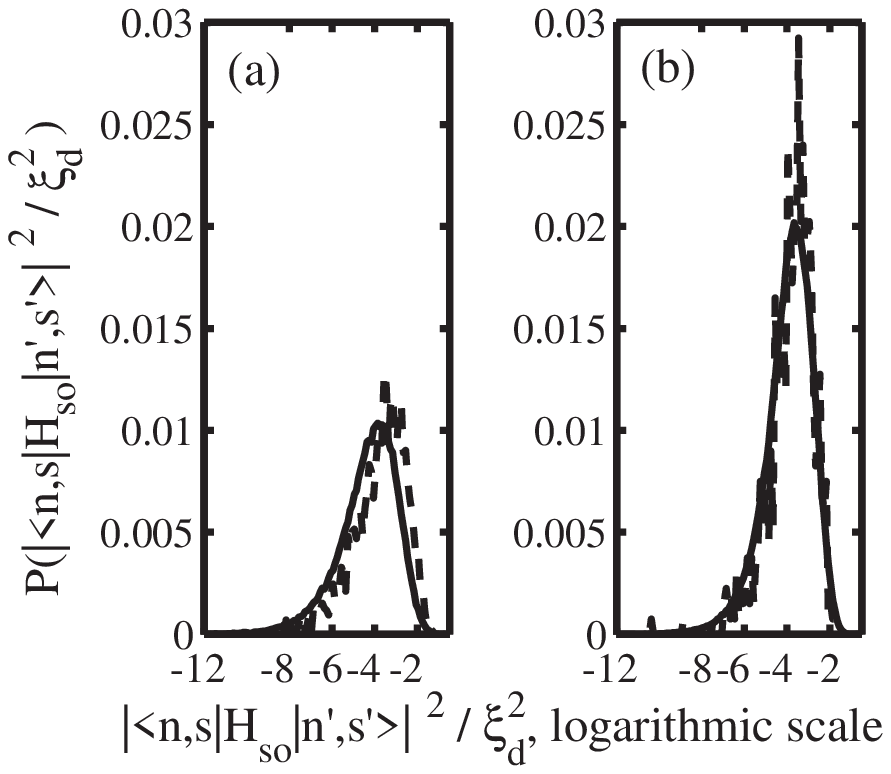}%
 \caption{Distribution function of spin-orbit matrix elements
for a spin-polarized 143-atom nanoparticle. $|n,s\rangle$ and $|n',s'\rangle$
are single-particle levels in absence of spin-orbit interaction. 
(a) $|n,s\rangle$
and $|n',s\rangle$ have like-spins; (b)
$|n,s\rangle$ and $|n,-s\rangle$ have opposite-spins.
Solid and dotted lines are for the cases of nearby levels
($|\epsilon^0_{n,s} -\epsilon^0_{n',s'}|/\delta <3$) 
and {\it any} pair of levels
respectively.}
 \label{hist4pub}
 \end{figure}
The typical size of avoided crossing gaps between opposite spin orbitals
in small nanoparticles can be understood by the following argument.  We first note 
that the unperturbed
orbitals satisfy the following sum rule
\begin{equation}
\sum_{n'} |\langle n',\downarrow| \vec L \cdot \vec S | n,\uparrow \rangle |^2 =
\langle n |L_{-} L_{+} |n \rangle)/4  \sim 2/3
\label{sum_rule}
\end{equation}
The estimate for the right hand side of Eq.~\ref{sum_rule} is based on the
on the same considerations leading to Eq.~\ref{eq:sumrule}, and uses that
$L_{-} L_{+} \sim 2/3 \vec L \cdot \vec L \sim 8/3$. 
If angular momentum matrix elements between orbitals
are not correlated with the energy differences between these orbitals, and the
matrix elements are reasonably narrowly distributed, this sum rule can be used
to estimate the typical matrix element.  
Expectation values of the angular momentum operators are zero because of 
angular momentum quenching, and a finite fraction of the matrix elements
vanish because of symmetries present in our rather regularly shaped 
nanoparticles.  Aside from these features, 
we find numerically that correlations between matrix elements and
energy differences are too small to be clearly observable. 
Fig.~\ref{hist4pub} shows the distribution function we have
obtained for matrix elements between opposite and like-spin states.
We have considered both the
matrix element distribution for closely spaced levels and 
the distribution for any pair of levels, not necessarily nearby.
The distributions are found to be very similar. Approximately 
$50\%$ and $70\%$ of matrix elements are zero for opposite-spin 
and like-spin cases respectively. 

Based on these numerical results and the sum rule Eq.~\ref{sum_rule}, 
we estimate the
typical value of $|\langle n',\downarrow| H_{\rm SO}| n,\uparrow \rangle |^2$
as ${2\over 3}\xi_d^2$ divided by half the total number of s, p and d orbitals, 
$9 {\cal N}_{a}/2$.
This implies a typical non-zero matrix element equal to $ \sim \xi_d \sqrt{.14/{\cal N}_{a}}$.

\begin{figure}
 \includegraphics[width=11.cm,height=11.cm]{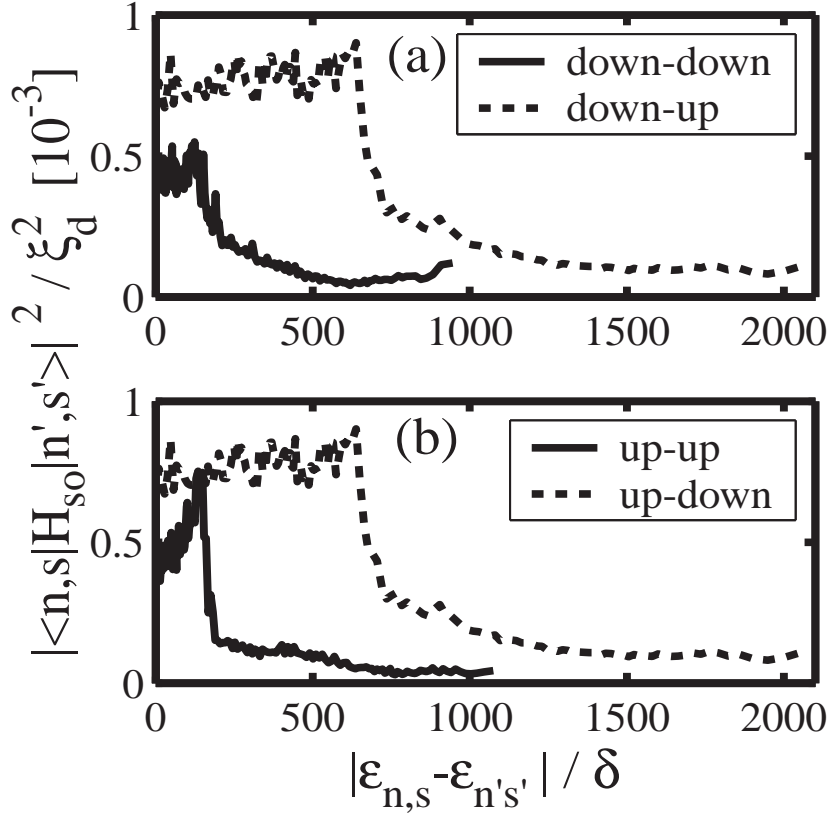}%
 \caption{Average squared matrix elements of the spin-orbit interaction
vs. energy difference for a polarized 143-atom nanoparticle.
$|n,s\rangle$ and $|n',s'\rangle$ are quasi-particle states without spin-orbit
interactions. The four curves correspond to the four possible spin
combinations.}
 \label{dist4}
 \end{figure}

In Fig.~\ref{dist4} we plot the average square matrix element 
vs. energy difference
for a 143 atom cluster, obtaining remarkably precise 
agreement with this estimate
provided that the energy differences are much smaller than the 
$d$-band width.

The average square matrix element for like spin orbitals is approximately 
a factor of
two smaller, consistent with the type of argument presented above which would
imply proportionality to $\langle n |L_{z}^2  |n \rangle$ in that case.

We expect the dependence of quasiparticle energies on the magnitude
and orientation of the order parameter, and also on external fields discussed
below, to change in character when the typical matrix element 
becomes comparable to
the level spacing, {\it i.e.} when

\begin{equation}
\frac{0.14 \xi_d^2}{\delta {\cal N}_a} = \delta\; .
\end{equation}

Our numerical calculations of quasiparticle spectra are consistent with using the 
condition $\hbar \tau_{SO}^{-1} \sim \epsilon_{SO} = \delta$ as a criterion for the 
{\it start} of the crossover to the strong coupling limit, and the condition
that the typical avoided crossing gap estimate be equal 
to $\delta$, as a criterion for {\it completion} of the crossover 
to strong-coupling relatively rigid spectra.  

For Cobalt nanoparticles the later condition is reached
for ${\cal N}_{a} \sim 2000$; for smaller nanoparticles, the quasiparticles 
generally have somewhat  
well-defined spin character, some Poisson character 
in their spectral statistics,
and complicated evolution patterns with external field 
and order parameter variations.
For larger nanoparticles, which we are not however able to study numerically,
we expect that quasiparticles will have strongly mixed spins, and more rigid spectra
with smoother evolution patterns.  
All the ferromagnetic nanoparticles that we are
able to study here, and many nanoparticles studied experimentally, are in the
crossover regime.  Note that since $\delta \propto W_d/{\cal N}_{a}$, 
$ (\xi_d^2/(\delta {\cal N}_a) \propto \hbar \tau_{SO}^{-1}$; the two 
conditions differ quantitatively not parametrically.  

 \begin{figure}
 \includegraphics[width=11.cm,height=11.cm]{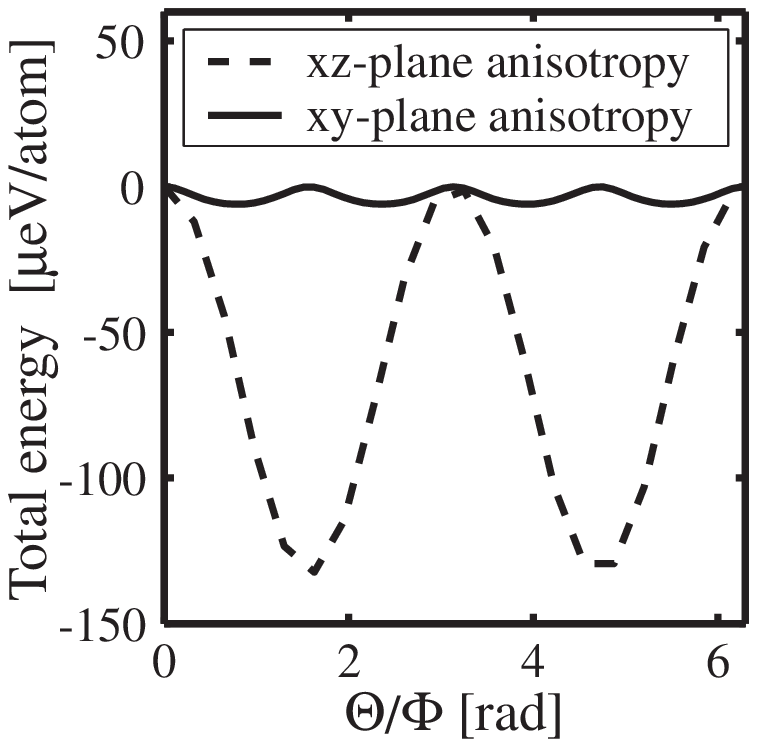}%
 \caption{Magnetic anisotropy energies 
for a 143-atom hemispherical nanoparticle. The dotted line
represents
$\big[E(h^{\star}_{\hat z}\,\hat z) - 
E(h^{\star}_{\hat \Omega}\,{\hat \Omega})\big]$
vs. the angle $\Theta$ between $\hat \Omega$
and $\hat z$, when $\hat \Omega$ is in the $zx$-plane.
The black line represents 
$\big[E(h^{\star}_{\hat x}\,\hat x) - 
E(h^{\star}_{\hat \Omega}\,{\hat \Omega})\big]$ vs.
the angle $\Phi$
between $\hat \Omega$ and $\hat x$, when $\hat \Omega$ lies in the $xy$-plane.}
 \label{anis2}
 \end{figure}

The total ground-state energy of the particle 
$E\big(h^{\star}_{\hat \Omega}\,\hat \Omega\big)$, obtained by summing 
over the lowest $N$ occupied orbitals, depends on the
direction of the self-consistent spin-splitting field $\hat \Omega$.
In Fig.~[\ref{anis2}] we plot 
$E(h^{\star}_{\hat z}\,\hat z) - E(h^{\star}_{\hat \Omega}\,{\hat \Omega})$
vs. the angle $\Theta$ between ${\hat \Omega}$ and
$\hat z$ (when ${\hat \Omega}$ lies in the $zx$-plane) and 
$E(h^{\star}_{\hat x}\,\hat x) - E(h^{\star}_{\hat \Omega}\,{\hat \Omega})$
vs. the angle $\Phi$ between  ${\hat \Omega}$ and 
$\hat x$ (when ${\hat \Omega}$ lies in the $zx$-plane).
From the figure we can see that the $xy$-plane is 
almost an {\it easy} plane for the model nanoparticle, except for a weak energy
dependence which generates four easy axis in the directions
$(\pm \hat x \pm \hat y)/2$. Again this property reflects the large significance of the 
overall sample shape.  
The four easy axis directions are remnants of the
magnetic anisotropy symmetry in bulk f.c.c. ferromagnets.  
Group theory considerations demand that a bulk f.c.c. ferromagnet 
have an easy axis in one of the directions perpendicular to the 8 
$(\pm 1, \pm 1, \pm 1)$\cite{skomski} planes. 
In a hemispherical nanoparticle the f.c.c. symmetry
is partially lifted 
and the magnetization is forced to lie in the $xy$-plane.
We can define the following 
two anisotropy-energy-per-atom constants
\begin{eqnarray}
k(\hat z,\hat x) \equiv {E(h^{\star}_{\hat z}\,\hat z) -
E(h^{\star}_{\hat x}\,\hat x)\over {\cal N}_a}
\approx 0.13 {\rm meV}\; ,
\label{anisconst1}\\
k(\hat x,\hat y) \equiv 
{E( h^{\star}_{\hat x}\,\hat x) -
E(h^{\star}_{\hat y}\,\hat y)\over {\cal N}_a}
\approx 0.01 {\rm meV}\; .
\label{anisconst2}
\end{eqnarray}
As expected on the basis of the 
qualitative considerations of Section~\ref{perturb},
the anisotropy per occupied orbital is much smaller than the average single-particle
level shift $\epsilon _{\rm SO}$ (approximately 200 times smaller in both cases) 
due to cancellations between positive and negative shifts mentioned above.

We notice that $k(\hat z,\hat x)$ for our nanoparticle is {\it larger}
than the bulk anisotropy per atom $k_{\rm bulk}=60 \mu$eV, 
as one would expect because of 
the hemispherical shape of the sample.  
However $k(\hat x,\hat y)$ is actually {\it smaller} 
than $k_{\rm bulk}$. This comparison should be regarded with caution,
since it is known that accurate theoretical estimates of the magnetic
anisotropy for bulk crystals are very delicate  
and agreement with experiment values even in the bulk is still not completely 
satisfactory \cite{zangwill2001};

we have not evaluated the anisotropy energy that 
results from the bulk limit of our nanoparticle model and it may well
not agree with experiment.  
Nonetheless, the small value that we find for the anisotropy in the $xy$-plane
might be connected
to the puzzling finding, in both tunneling\cite{deshmukh2001} and switching-field\cite{jamet2001}
experiments on single ferromagnetic
nanoparticles, of anisotropy energies per atom a factor of five
{\it smaller} than bulk values.

\section{Fluctuations of magnetic anisotropy}
\label{fluct}

So far we have considered only average values of the anisotropy constants.
In ferromagnetic nanoparticles, however, these quantities are also characterized by large
fluctuations as a function of experimentally relevant parameters. Anisotropy fluctuations are
the topic of this section.
\subsection{Charge-induced fluctuations 
as a function of the spin-splitting field}

 \begin{figure}
 \includegraphics[width=16.cm,height=8.cm]{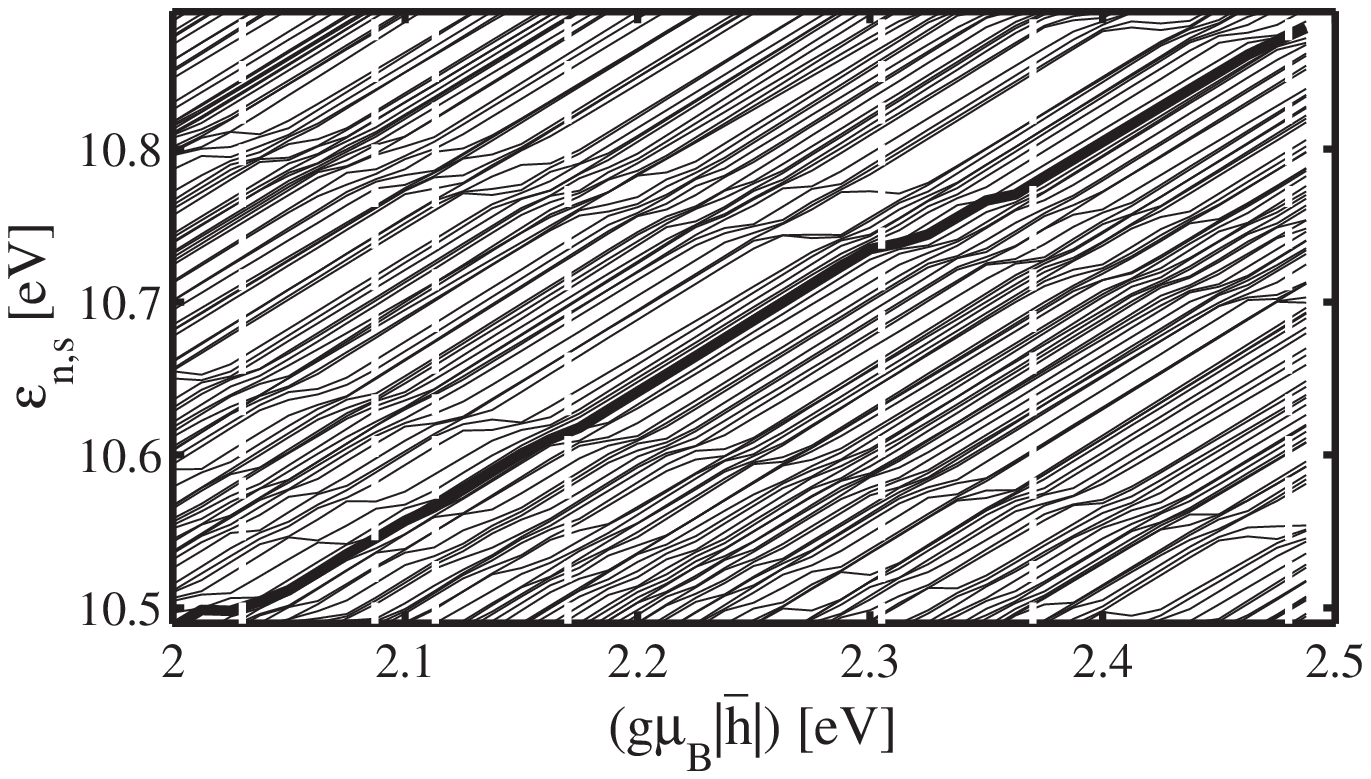}%
\caption{Change of quasiparticle energies with increasing spin-splitting
field $|\vec h|$
near $h^{\star}$ for a 143-atom nanoparticle.
200 levels are shown.
The Fermi level is the thick black line. The vertical white dashed lines 
indicate the positions 
of the charge redistributions displayed in Fig.~[\ref{anisfluc143}].
The vertical black dashed line corresponds to the self-consistent
spin-splitting field.}
 \label{eigspectrum2}
 \end{figure}

 \begin{figure}
 \includegraphics[width=11.cm,height=11.cm]{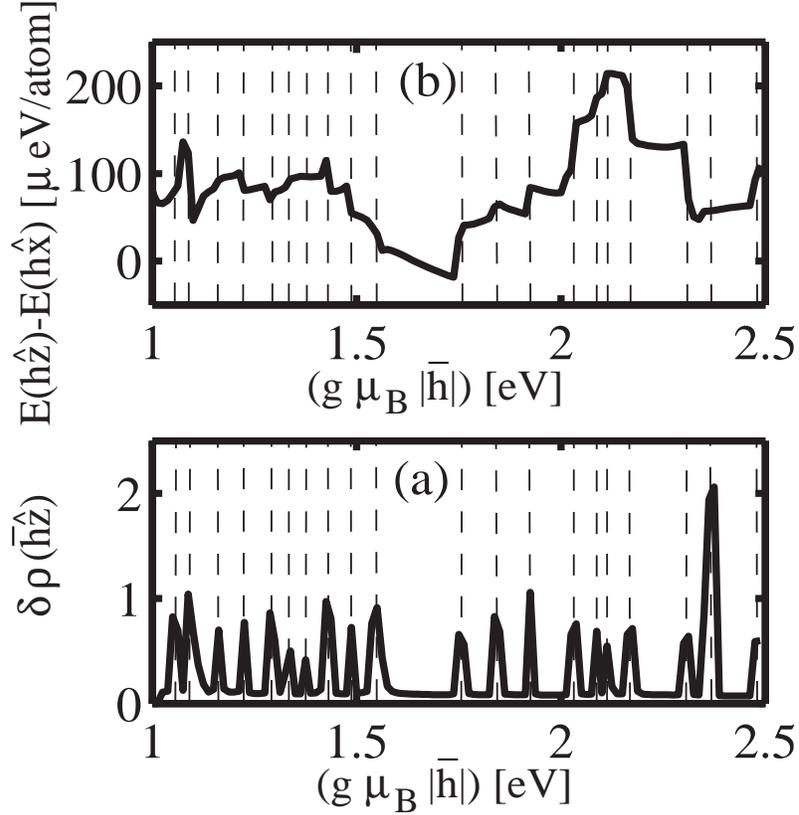}%
 \caption{Relation between charge fluctuations and anisotropy 
in 143-atom nanoparticle. 
(a) Total charge fluctuation defined in Eq.~[\ref{charge_fluct}]. 
Charge
fluctuations of the order of one electron result in fluctuations in the
magnetic anisotropy.
(b) Anisotropy energy in the $zx$-plane vs.
the magnitude of the spin-splitting field taken as a free parameter.}
 \label{anisfluc143}
 \end{figure}

When spin-orbit interactions are included, 
all quasi-particle eigenstates have mixed 
majority-spin and  minority-spin character. 

Our specific calculations, however, are for small enough particles such that
most eigenstates still have predominantly one spin 
character. It is convenient to use this ``predominant spin'' to label 
the states when discussing their dependence on $|\vec h|$  or on an external 
magnetic field. In a paramagnetic system ($\vec h^{\star}=0$) in absence
of external field, there is a Kramer degeneracy that pairs up eigenstates
with opposite predominant spin character. 
The degeneracy is lifted in the ferromagnetic state.  Majority-spin
states will move down in energy while minority-spin states will move up
as $|\vec h|$ increases. Because of spin-orbit--induced
level repulsion, all of the level crossings are avoided; 
the levels evolve continuously with $|\vec h|$, gradually changing
their spin and orbital character. 
In Fig.~[\ref{eigspectrum2}] we plot
the variation of two hundred levels as a function
of $g_s\mu_{\rm B}|\vec h|$ near $g_s\mu_{\rm B}h^{\star}=2.2$ eV,
for a 143-atom nanoparticle. The Fermi level is the thick black line
lying in a region of predominantly minority-spin quasiparticles
(lines with positive slope). By increasing $|\vec h|$, individual
majority-spin quasiparticle energies (negative slopes) 
come down from regions above
the Fermi level, creating avoided crossing gaps as they approach
minority-spin quasiparticles moving in the opposite direction. 
Since a 143-atom nanoparticle is still toward the {\it weak}
spin-orbit coupling limit, the level crossings will be avoided
only weakly.
Whenever one of the majority-spin quasiparticles crosses the Fermi level,
there is a change in the spin character of one of the quasiparticle levels and
the total spin of the nanoparticle increases approximately by one.
Due to the exchange interaction, such spin-flips bring about  
charge redistribution inside the nanoparticle, which in turns
gives rise to fluctuations in the anisotropy energy. 
In Fig.~[\ref{anisfluc143}a] we
plot the total atomic site charge redistribution,
when $h \equiv |\vec h|  \Rightarrow h + \Delta h$, for $\Delta h = 12.5$ meV
\begin{equation}
\delta \rho(h) = \sum_i
\Big|\rho_i(h +\Delta h)-\rho_i(h)\Big| 
\approx \sum_i \Bigg| {d\rho_i(h)\over dh}\Bigg|\Delta h\; ,  
\label{charge_fluct}
\end{equation}

where $\rho_i(h)$ is the total charge at atom $i$. It is seen that
$\delta \rho(h)$ changes by one in the small interval $\Delta h= 12.5$ meV
where majority-spin quasiparticles weakly avoid crossing the Fermi level. 
(Note that this is a charge redistribution, not a change in total charge.) 
This result is not unexpected, since the majority and 
minority spin orbitals should have uncorrelated spatial distributions.
The charge redistribution is likely overstated by our simplified model,
however, since it does not account for long-range Coulomb interactions and 
related screening effects. It is, however still a relatively small 
$\sim 1/N$ effect.  The anisotropy energy fluctuations
associated with the level crossings that occur as a function of 
$h$, shown in Fig.~[\ref{anisfluc143}b] are much larger in relative terms.

Variation in $h$ corresponds to variation in the amplitude of the magnetic 
order parameter. In micromagnetic modeling of nanoparticles, it is implicitly
assumed that the magnitude of the order parameter is fixed and that this 
collective degree of freedom can be ignored in modeling nanoparticle properties. 
The substantial dependence of anisotropy energy on $h$ that we
find demonstrates that the amplitude and orientation fluctuations of the 
order parameter can be strongly coupled in small magnetic nanoparticles. 
Related anisotropy fluctuations as a function of 
{\it electron number} are important in understanding the addition-potential-spectroscopy
single-electron-transistor experiments\cite{deshmukh2001} which are
the topic of the next section. 

\subsection{Mesoscopic fluctuations as a function of electron number}
\label{anis_fluctsN}

The analysis of the field-dependence of tunneling resonances in
experiments on magnetic nanoparticles suggests that the anisotropy
energy fluctuates significantly from eigenstate 
to eigenstate\cite{deshmukh2001,kleff2001prl}. 
In Refs.~\cite{kleff2001prl,deshmukh2001} the effects of these
fluctuations were 
mimicked by using two different
anisotropy energy constants $k_{N}$ and 
$k_{N\pm 1}= k_{N} +\delta k_{\pm}$ for $N$- and $N\pm 1$-electron states. 
By assuming $\delta k_{\pm}/k_{N}$ in the range of a few percent, it
was possible to explain the non-monotonic behavior of the tunneling
resonances seen experimentally.

Within our microscopic model,
we have calculated $k_{N}$ and $k_{N\pm 1}$ as defined in 
Eqs.~(\ref{anisconst1},\ref{anisconst2}).
In light of results of the type summarized in Fig.~[\ref{anis2}],
it suffices to compute the
total energy for two different directions of the spin-splitting field
and take the difference.
The total energy for a $N$-electron system,
for spin splitting fields along directions 
$\hat \Omega_i\;,\ i=1,2$, is given by:

\begin{equation}
E({h}^{\star}_i\,\hat \Omega^{\phantom {\star}}_i)=
\sum_{n=1}^{N}
\epsilon_{n}({h}^{\star}_i\, \hat\Omega^{\phantom {\star}}_i) +
\frac{({h}_i^{\star})^{2}}{2U_{dd}}{\cal N}_a\;,\ \ \ i =1,2\;,
\end{equation}
where ${h}^{\star}_i$ is calculated self-consistently
for a given fixed direction. The anisotropy-energy-per-atom constant is then
\begin{equation}
k_N(\hat \Omega_1, \hat \Omega_2) 
= {1\over {\cal N}_a}\Big[
\sum_{n=1}^{N} \epsilon_{n}({h}^{\star}_1\, \hat\Omega^{\phantom {\star}}_1) -
\epsilon_{n}({h}^{\star}_2\, \hat\Omega^{\phantom {\star}}_2)\Big]
+ \frac{({h}_1^{\star})^{2}-({h}_2^{\star})^{2}}{2U_{dd}} (g_s \mu_{\rm B})^2
\end{equation}
where the subscript $N$ emphasizes the fact that the constant refers to a 
$N$-electron system.
It turns out that the value of ${h}^{\star}$ depends 
very weakly on $\hat\Omega$. For example, in a 143-atom nanoparticle 
$|{h}^{\star}_1 - {h}^{\star}_2|/{h}^{\star}_1 \approx 10^{-3}$. 
This property reflects the large ratio between the total magnetic condensation energy 
and the anisotropy energy, even for nanoparticles.  The strong coupling
between amplitude and orientation fluctuations mentioned above, occurs only
when a small change in $h$ leads to a crossing between majority and minority 
spin orbitals.  Evaluating $E({h}\,\hat \Omega^{\phantom {\star}}_2)$
at $h= {h}^{\star}_1 = {h}^{\star}_2 + \delta {h}^{\star}$, expanding in
powers of $\delta {h}^{\star}/{h}^{\star}_1$ and remembering that
${h}^{\star}_2$ minimizes $E({h}\,\hat \Omega^{\phantom {\star}}_2)$
yields
\begin{equation}
k_N(\hat\Omega_1,\hat \Omega_2) = 
{1\over {\cal N}_a}\Big[\sum_{n=1}^{N} 
\epsilon_{n}({h}^{\star}_1\, \hat\Omega^{\phantom {\star}}_1) -
\epsilon_{n}({h}^{\star}_1\, \hat\Omega^{\phantom {\star}}_2)\Big]
+ {\cal O}\Bigg[{N\over {\cal N}_a}
\Big({\delta {h}^{\star}\over {h}^{\star}_1}\Big)^2
\Bigg]
\end{equation}
for any $N$.

 \begin{figure}
 \includegraphics[width=11.cm,height=11.cm]{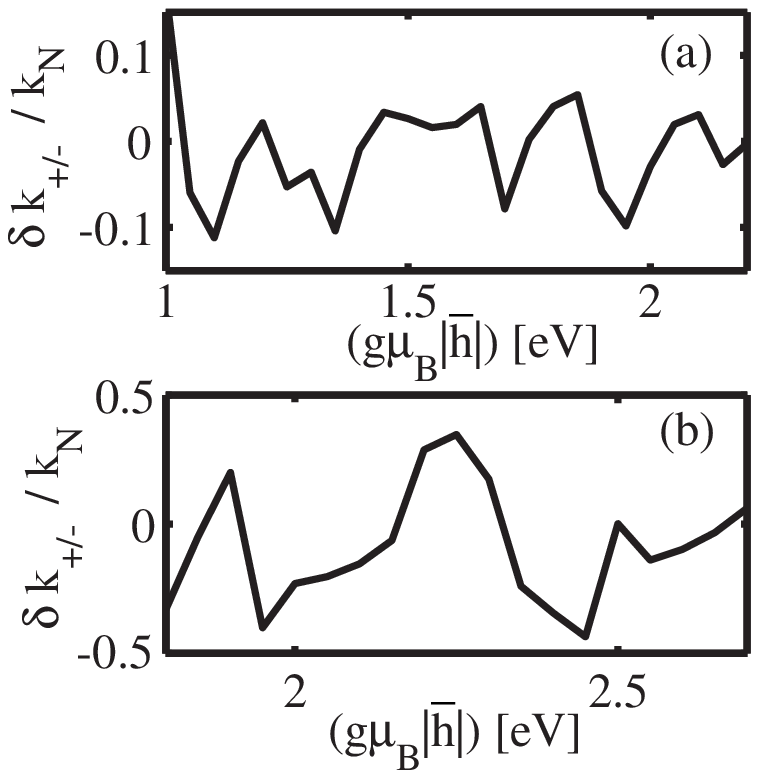}%
 \caption{Relative shift $\delta k_{\pm}/  k_N \equiv (k_{N+1}-k_N)/k_N$ 
of the anisotropy energy constant 
$k_N(\hat z,\hat x)$, 
when the electron number increases by one, $N \to N+1$,
as a function of spin-splitting field. (a) 260-atom nanoparticle. (b) 143-atom
nanoparticle. The self-consistent spin-splitting field is 
$h^{\star}\approx  2.2$ eV for both a 143-atom and a 260-atom nanoparticle.}
 \label{aniscomp}
 \end{figure}
Adding an electron to the system changes the magnitude of the
spin-splitting field ${h}^{\star}_{i} \to {h}^{\star,}_{i}$, but
again $|{h}^{\star,}_{i} - {h}^{\star}_{i}|/{h}^{\star}_{i} <<1$. 
Thus we obtain
\begin{eqnarray}
k_{N+1}(\hat \Omega_1, \hat \Omega_2) =
{1\over {\cal N}_a}\Big[\sum_{n=1}^{N+1}
\epsilon_{n}({h}^{\star}_1\, \hat\Omega^{\phantom {\star}}_1) -
\epsilon_{n}({h}^{\star}_1\, \hat\Omega^{\phantom {\star}}_2)\Big]
+ {\cal O}\Bigg[{N+1\over {{\cal N}_a}}
\Big({\delta {h}^{\star}\over {h}^{\star}_1}\Big)^2\Bigg]\nonumber\\
= k_N(\hat \Omega_1, \hat \Omega_2) + 
{\epsilon_{N+1}({h}^{\star}_1\, \hat\Omega^{\phantom {\star}}_1) -
\epsilon_{N+1}({h}^{\star}_1\, \hat\Omega^{\phantom {\star}}_2)
\over {\cal N}_a} +
{\cal O}\Bigg[{N+1\over {{\cal N}_a}}
\Big({\delta {h}^{\star}\over {h}^{\star}_1}\Big)^2\Bigg]
\end{eqnarray}
Replacing 
\begin{equation}
{\epsilon_{N+1}({h}^{\star}_1\, \hat\Omega^{\phantom {\star}}_1) -
\epsilon_{N+1}({h}^{\star}_1\, \hat\Omega^{\phantom {\star}}_2)
\over {\cal N}_a}
\approx {\Delta[\delta \epsilon_{N+1}(\hat\Omega_1, \hat\Omega_2)]\over {\cal N}_a}\; ,
\end{equation}
where $\Delta[\delta \epsilon_{N+1}(\hat\Omega_1, \hat\Omega_2)]$
is the width of the distribution of the single-particle anisotropies in plane containing the two directions 
$\hat\Omega_1$ and $\hat\Omega_2$,
we finally obtain
\begin{equation}
k_{N+1}(\hat \Omega_1,\hat \Omega_2)
\approx k_N(\hat\Omega_1,\hat\Omega_2) + 
{\Delta[\delta \epsilon_{N+1}(\hat\Omega_1, \hat\Omega_2)]\over {{\cal N}_a}}\; ,
\end{equation}
The fluctuations of $k_N$ due to an additional electron are therefore 
of the order of $\Delta[\delta \epsilon_{N+1}(\hat\Omega_1, \hat\Omega_2)]/{\cal N}_a$,
and are regulated 
by the mean level spacing, $\delta$, which suppresses the magnitude of 
$\Delta[\delta \epsilon_{N+1}(\hat\Omega_1, \hat\Omega_2)]$ 
at large nanoparticle sizes as 
we discuss below.  

For a 143-atom dot, $\Delta[\delta \epsilon_{N+1}(\hat\Omega_1, \hat\Omega_2)]
\approx 2.9 $meV in the zx-plane, where $k_N(\hat z, \hat x) \approx 0.13$ meV.
Therefore $\delta k_{\pm}/k_N(\hat z, \hat x)=
{1\over {\cal N}_a}\Delta[\delta \epsilon_{N+1}(\hat\Omega_1, \hat\Omega_2)]
/k_N(\hat z, \hat x)\approx 15\%$.
These estimates are confirmed by a direct calculation of 
$k_N(\hat z, \hat x)$ and $k_{N+1}(\hat z, \hat x)$ for  143-atom
and  260-atom nanoparticles, shown in Fig.~[\ref{aniscomp}]. Here
$k_N(\hat z, \hat x)$ and $k_{N+1}(\hat z, \hat x)$ are plotted as a
function of $g_s\mu_{\rm B} h$, taken as a free parameter 
(the self-consistent value is close to
$g_s\mu_{\rm B} h^{\star} = 2.2$ eV for both a 143-atom and a 260-atom
nanoparticle).
For a 143-atom nanoparticle the calculated anisotropy energy fluctuations are
of the order of $20\%$.
For a 260-atom nanoparticle the anisotropy energy fluctuations
are smaller, but still close to $5\%$.

We can understand the surprisingly large fluctuations in anisotropy energy with
particle and electron number by the following observations.
Our numerical
results (see Fig.~\ref{eiganis3} and Fig.~\ref{histeiganis2}) 
indicate that the contribution of a 
given orbital to the anisotropy energy
is chosen essentially at random from a distribution which, for the $zx$-plane
anisotropy, has a width
$\Delta(\delta\epsilon_{n,s})\sim 2.2\,\xi_d^2 / W_d \sim 2.9$ meV and a much smaller mean value 
$ \langle \delta\epsilon_{n,s}\rangle = 
\langle \epsilon_{n,s}(h^{\star}\hat z) -
\epsilon_{n,s}(h^{\star}\hat x)\rangle = 
{\cal N}_a k_N(\hat z,\hat x)/N \sim 15 \mu$eV,
where
$k_{N}(\hat z,\hat x) \sim 0.13$ meV. 
The relative change in the anisotropy energy expected when
one additional orbital is occupied is 
$\sim 4.3\,\xi_d^2/ W_d/ (k_{N}(\hat z,\hat x)  {\cal N}_{a})$,
which can easily be larger than $1\%$ for ${\cal N}_{a} \le  1500$.

We expect strong level repulsion in the rigid spectrum of larger nanoparticles
to be accompanied by more regular behavior of the anisotropy
energy per atom, with smaller variations as a function 
$\vec h^{\star}$, and electron number.  

We note that fluctuations in the contribution of a given orbital to
the anisotropy energy should not exceed $\sim \delta$, which vanishes 
in the limit of very large particles.  We expect that the distribution function of 
contributions to anisotropy from orbitals near the Fermi energy to become
narrow when $\delta $ is smaller than the average anisotropy energy contribution
$\sum 10 {\rm \mu eV}$.  This condition is satisfied for particles 
containing more than $\sim 10^{5}$ atoms. 

\subsection{Anisotropy energy dependence on particle atom number and shape}

 \begin{figure}
 \includegraphics[width=11.cm,height=11.cm]{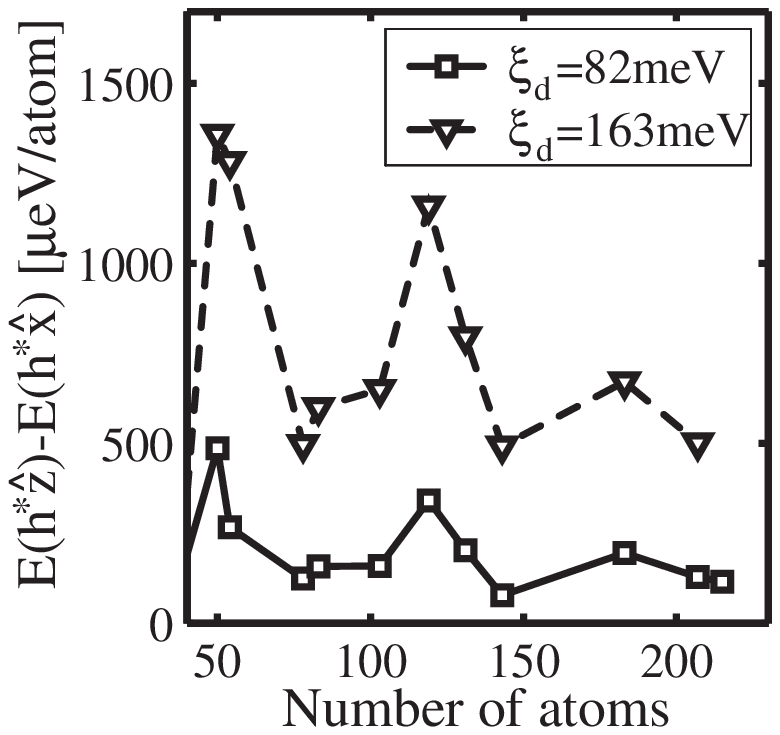}%
 \caption{Anisotropy energy as a function of nanoparticle size for two values of the
spin-orbit coupling $\xi_d$. A splitting
field of $2$ eV is assumed for all sizes.}
 \label{anisatom2}
 \end{figure}

We conclude this section with a few remarks on the dependence
of the magnetic anisotropy energy on nanoparticle size and shape.
In Fig.~[\ref{anisatom2}] we plot the anisotropy energy per atom
for nanoparticles of different sizes.
For a small number of atoms (below 60) the anisotropy per atom is very 
large and decreases rapidly with size.
In this regime, non-extensive surface contributions, which are 
present because of the abrupt truncation of the lattice,
are clearly playing a dominating role. 
By increasing the size of the nanoparticle,
we must eventually reach a regime where  
total magnetic anisotropy becomes proportional to the nanoparticle volume.
Fig.~[\ref{anisatom2}] shows that this regime is not yet  
fully established even for nanoparticles with hundreds of atoms, 
and fluctuations are still pronounced. In this regime
the anisotropy per atom is still 2-3 times larger than the
bulk value for Cobalt, which is 0.06 meV.
There is no regime in which {\em bulk} (proportional to ${\cal N}_{a}$) and 
surface (proportional to ${\cal N}_{a}^{2/3}$) contributions to the 
anisotropy can be cleanly separated. 

 \begin{figure}
 \includegraphics[width=11.cm,height=11.cm]{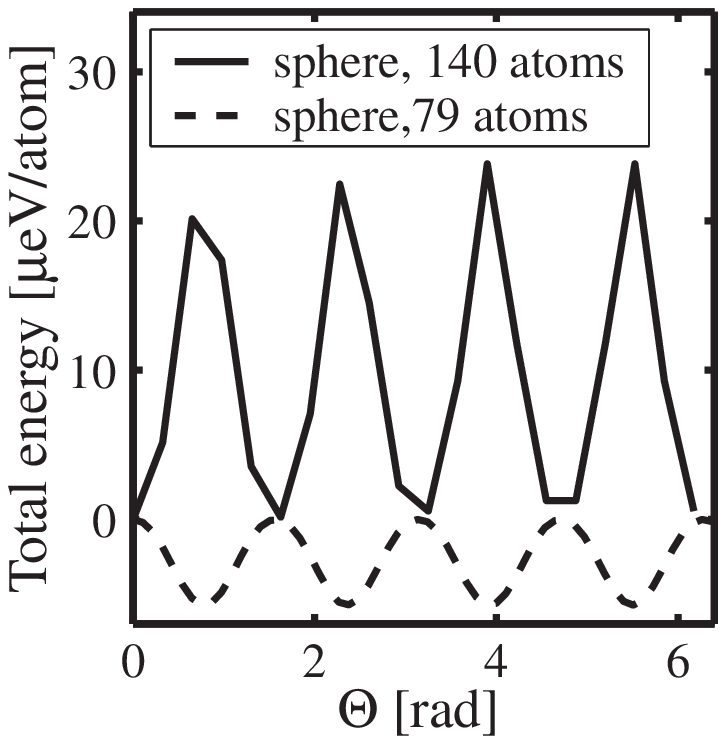}%
 \caption{Magnetic anisotropy energy 
in the $zx$-plane, 
$\big[E(h^{\star}_{\hat z}\,\hat z) -
E(h^{\star}_{\hat \Omega}\,{\hat \Omega})\big]$
for spherical nanoparticles
of two sizes. $\Theta$ is the angle between $\hat z$ and  $\hat \Omega$.
By symmetry, the anisotropy energy in the $xy$-plane
is now the same as in the
$zx$-plane.
  The spin-splitting field is assumed to be $2$ eV for both sizes.}
 \label{sphereanis}
 \end{figure}

In is clear that for small particles the shape of the nanoparticle plays an 
important role in the determination of 
the magnetic anisotropy.  For example, as seen in Fig.~[\ref{sphereanis}],
for spherical particles we find
anisotropy energies which are one 
order of magnitude smaller than the $zx$ anisotropy
of hemispherical particles, comparable instead to 
the $xy$ anisotropy of hemispherical particles
with the same number of atoms. A 140-atom spherical particle
has an anisotropy energy per atom  $\approx 0.01-0.02$meV. 

It is interesting to notice that 
both tunneling\cite{deshmukh2001} and switching-field\cite{jamet2001}
experiments on single ferromagnetic
nanoparticles find anisotropy energies per atom which are of the order
$0.01$meV, a factor of five
{\it smaller} than the bulk value. The nanoparticles in tunneling
experiments are roughly hemispherical\cite{gueron1999,deshmukh2001}, 
whereas the nanoparticle shape
in Ref.~[\onlinecite{jamet2001}] is close to spherical.

Further theoretical studies that focus on the relationship
between nanoparticle shape and anisotropy could be helpful to 
efforts to engineer ferromagnetic nanoparticles
whose shape, size, and crystal structure are tuned to produce 
desired magnetic properties.

\section{Hysteresis and variation of single-particle levels in an
external magnetic field}
\label{hysteresis}
 \begin{figure}
 \includegraphics[width=11.cm,height=11.cm]{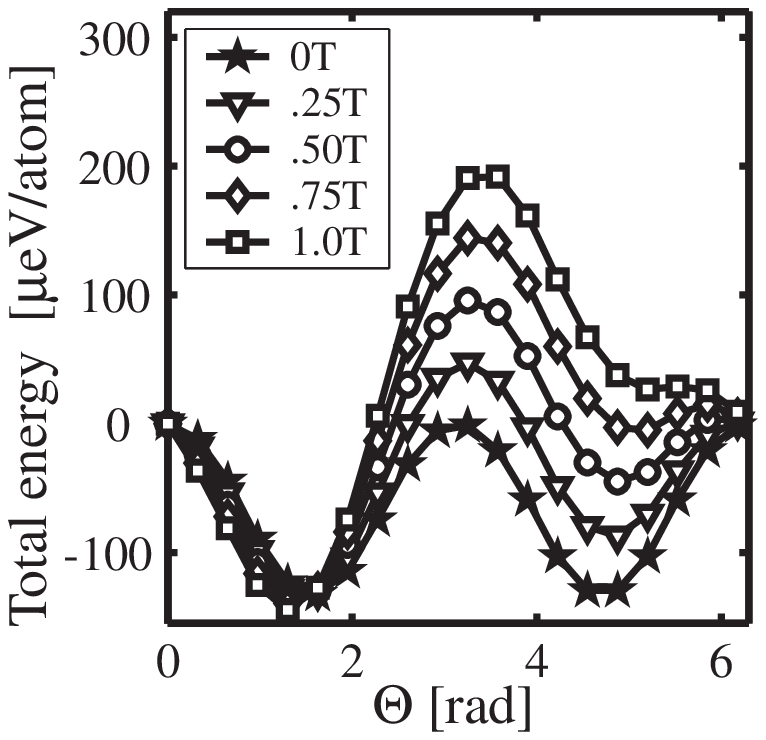}%
 \caption{``Uniaxial'' anisotropy energy as a function of the
direction of the spin-splitting field $\hat \Omega$, for different magnitudes
of the external magnetic field. The external field is in the 
$zx$-plane at an angle $\theta = \pi/4$ with the $\hat z$-axis.
$\Theta$ is angle between $\hat \Omega$ (lying in the $zx$-plane)
and $\hat z$.}
 \label{anismag}
 \end{figure}

 \begin{figure}
 \includegraphics[width=11.cm,height=11.cm]{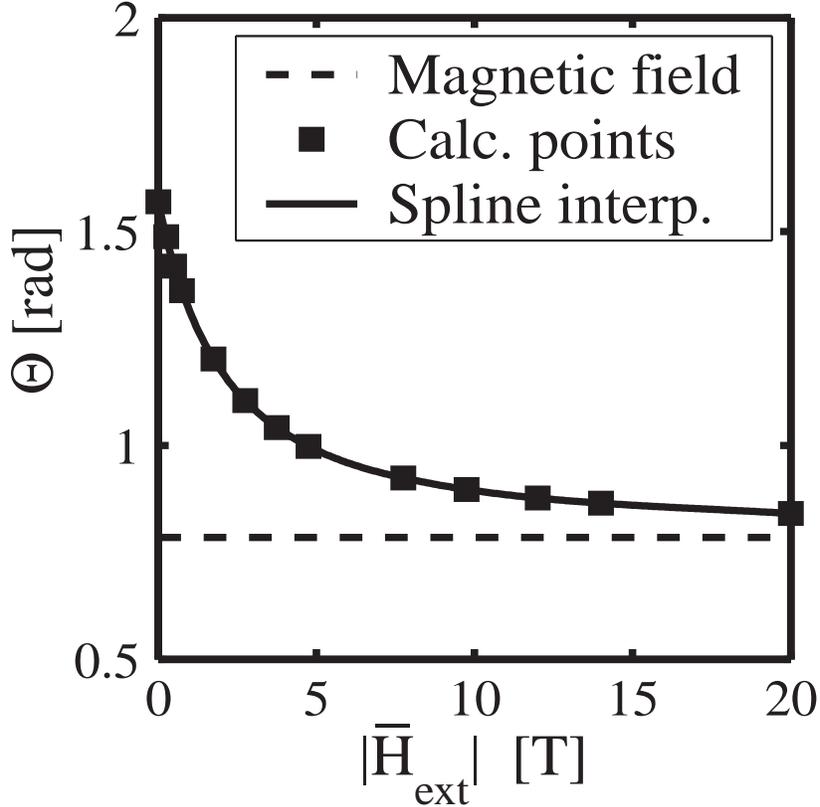}%
 \caption{Variation of the magnetization direction of the
stable minimum of Fig.~[\ref{anismag}], as a function of $|\vec H_{\rm ext}|$.
$\Theta$ is the angle between $\hat \Omega$ and $\hat z$. 
$\vec H_{\rm ext}$ is in the $zx$-plane at $\pi/4$ from the $\hat z$- axis.}
 \label{angle1}
 \end{figure}
 \begin{figure}
 \includegraphics[width=11.cm,height=11.cm]{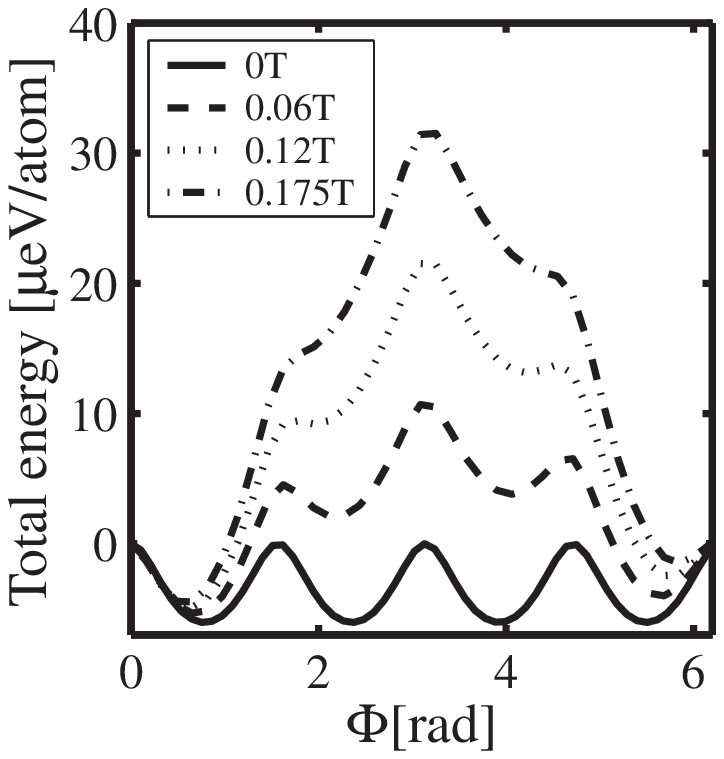}%
 \caption{Anisotropy energy in the $xy$-plane for different magnitudes
of an external magnetic field. The direction of the external field
is $(\theta= \pi/4, \phi= \pi/16)$. $\Phi$ is the angle between
$\hat \Omega$ (lying in the $xy$-plane) and $\hat x$.}
 \label{xyanisdb}
 \end{figure}

In the last section we shall investigate the effect of an external magnetic
field on the total energy of the nanoparticle. This will allow us
to make a connection between our microscopic model and more familiar
classical micromagnetic energy functional expressions
which are normally used to interpret the results of 
switching-field experiments\cite{jamet2001}.  We shall also study the
dependence of the quasiparticle energy levels on external magnetic fields;
this complex behavior has been probed directly in recent 
tunneling experiments\cite{gueron1999,deshmukh2001}.

\subsection{Hysteresis}
Ideally one would like to study the three dimensional energy landscape 
$E(\vec h^{\star}, \vec H_{\rm ext})$ as a function of  
$\hat\Omega(\Theta,\Phi) \equiv \vec h^{\star}/|\vec h^{\star}|$, for several
values of $\vec H_{\rm ext}$. Here we consider two simplified
cases. In Fig.~[\ref{anismag}] we plot the
total energy difference
$E(\vec h_{\hat z}^\star{\hat z})-E(\vec h_{\hat \Omega}^\star{\hat \Omega})$ 
as a function 
of an external magnetic field $\vec H_{\rm ext}$.
Here $\hat \Omega$ is allowed to rotate in the
$zx$-plane.
In this picture we recognize the familiar features
of a micromagnetic energy functional 
characterized by a uniaxial magnetic anisotropy:
at $\vec H_{\rm ext}=0$ there are two degenerate minima, separated by
an energy barrier, representing two equivalent easy directions $\pm \hat x$.
When a magnetic field is applied in the $zx$-plane, the degeneracy is lifted
and the minimum in the $-\hat x$ direction becomes a metastable local minimum, 
until a switching field is reached at which the local minimum disappears. 
This happens at $|\vec H_{\rm ext}| \approx 1$T for a 143-atom nanoparticle.
This simple model captures the
essence of classical hysteresis in ferromagnetic nanoparticles.
Quantum mechanically, for sufficiently weak fields the system can still sit
in the quantum state characterized by a magnetic moment pointing along
the classically metastable direction until the switching field is reached,
the system relaxes to its true ground state, and the magnetization orientation
collective coordinate changes discontinuously  
\footnote{\label{fnote6}
This scenario neglects the possibility
of quantum tunneling between the two local minima through the energy barrier
separating them. At low fields and for nanoparticles of 1-4 nanometers
in diameter, quantum tunneling is exponentially small.}.

With a further increase of the external field, the magnetization is gradually
twisted from the easy $x$-axis toward the direction  of the
magnetic field, as shown in Fig.~[\ref{angle1}]. At low fields ($ |\vec H_{\rm ext}|<  1$T), the direction
of the magnetization corresponding to this local minimum is very close to
the $xy$-plane.

Because our hemispherical nanoparticles have a nearly isotropic easy 
$xy$-plane, the magnetization will 
rotate in the $xy$-plane in response to {\it weak} applied external fields\footnote{\label{fnote7}
Clearly, if the external field is not in the $xy$-plane
the magnetization will not stay exactly in this easy plane.
However the displacement of the magnetization from easy to hard directions
is negligible at the weak fields ($ |\vec H_{\rm ext}|< 1 T$) 
(See Fig.~[\ref{angle1}] )
and it becomes relevant only at larger fields, when the
component of the magnetization in the $xy$-plane is essentially
frozen along the direction of the component of $|\vec H_{\rm ext}|$ in that
plane.}
This is exemplified in Fig.~[\ref{xyanisdb}], where we plot the
the total energy as a function of $|\vec H_{\rm ext}|$, when  
$\Omega$ lies in the $xy$-plane. $|\vec H_{\rm ext}|$ is oriented
in the direction $(\theta = \pi/4, \phi=\pi/16)$.
In the absence of the external field,
the ground state of the nanoparticle is four-fold degenerate, with the
degeneracy corresponding to
one of the four magnetization directions $\pm \hat x \pm \hat y$, 
associated to the four local minima
of Fig.~[\ref{xyanisdb}]. As the external field increases, three
of these minima will become classically metastable with small barriers
separating them from each other and the true ground state. 
In general the different local minima will lose their
metastability at different external field strengths.
In the case we consider,
the first switching field is reached at $ |\vec H_{\rm ext}| \approx 0.15$T,
when the minimum originally at $-\hat x  - \hat y$ disappears. 
If the system
starts out in this local minimum at zero external field,
at the switching field it will jump 
to the ground state with a corresponding rotation of the 
magnetization in the $\hat x  + \hat y$-direction. 
Since
everything takes place essentially within the $xy$-plane, 
where the typical anisotropy
energies are one order of magnitude smaller than in the $zx$-plane, the
scale of the coercivity is also much smaller than the one in the $zx$-plane.
More complicated hysteretic behaviors are also possible, in which,
by starting from a different minimum at zero external field,
the system jumps first from one metastable to another metastable state
and only at a second switching field reaches its true ground state.


\subsection{Dependence of quasiparticle levels on external fields}

\begin{figure}
 \includegraphics[width=11.cm,height=11.cm]{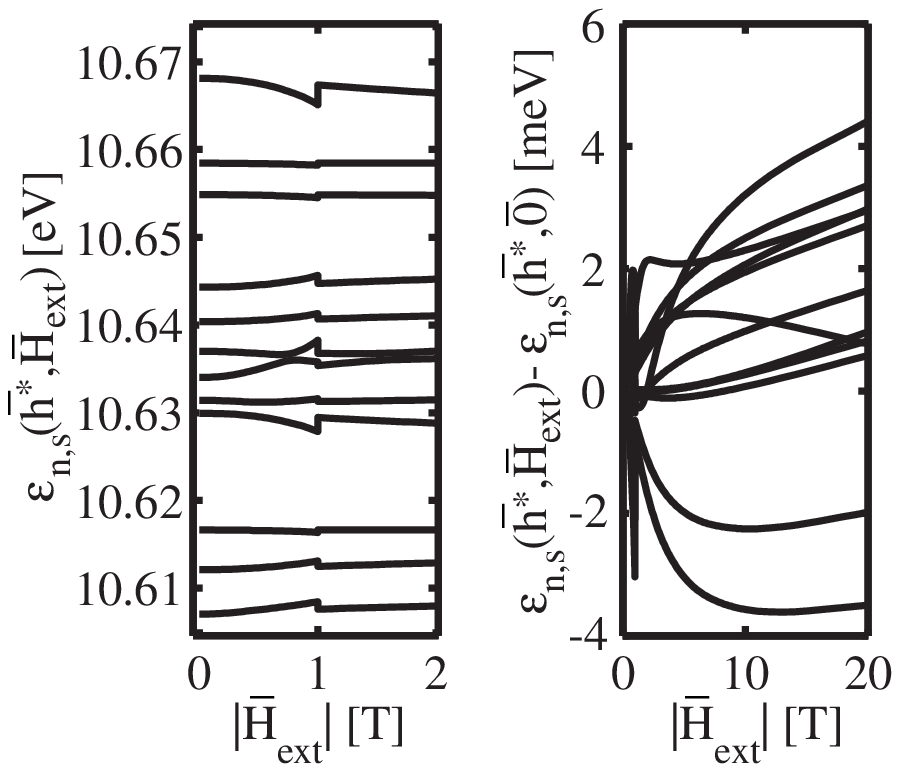}%
\caption{Variation of quasiparticle energy levels 
$\epsilon_{n,s}(\vec h^{\star}, \vec H_{\rm ext})$ 
around the Fermi level
in an external magnetic field.
The levels are calculated assuming that
the ground-state of the nanoparticle changes in the external field 
in the way described in Fig.~[\ref{anismag}].}
 \label{mod1}
 \end{figure}

The hysteretic behavior that we have seen in the ground-state properties 
of a ferromagnetic nanoparticle
has profound implications for the magnetic field dependence
of its low-energy elementary excitations. In ferromagnetic metals 
there exist two kinds of elementary excitations: collective 
spin excitations associated with magnetization orientation degrees
of freedom, and particle-hole excitations. In a ferromagnetic
nanoparticle the distinction between these two kinds of excitations
is partly obscured by the effect of spin-orbit 
coupling\cite{ahm_cmc2001ssc, ce_cmc_ahm2002pap4}.
Sorting out this problem is particularly important in order to
interpret current tunneling experiments in single-electron transistors\cite{deshmukh2001}.
Here we examine only particle-hole excitations
around the Fermi level, which are immediately available within
our Hartree-Fock treatment of the many-body Hamiltonian.
In Fig.~[\ref{mod1}] we plot the magnetic field dependence
of a few single-particle energy levels around the Fermi energy.
As a simplified illustration,
the levels are calculated assuming that the ground-state dependence
on the external field 
is as described in Fig.~[\ref{anismag}]: 
at low field the ground-state magnetization
is oriented around the $-\hat x$-axis until the switching field is 
reached, whereupon the magnetization is reversed along a direction
around the $\hat x$-axis.
 \begin{figure}
 \includegraphics[width=11.cm,height=11.cm]{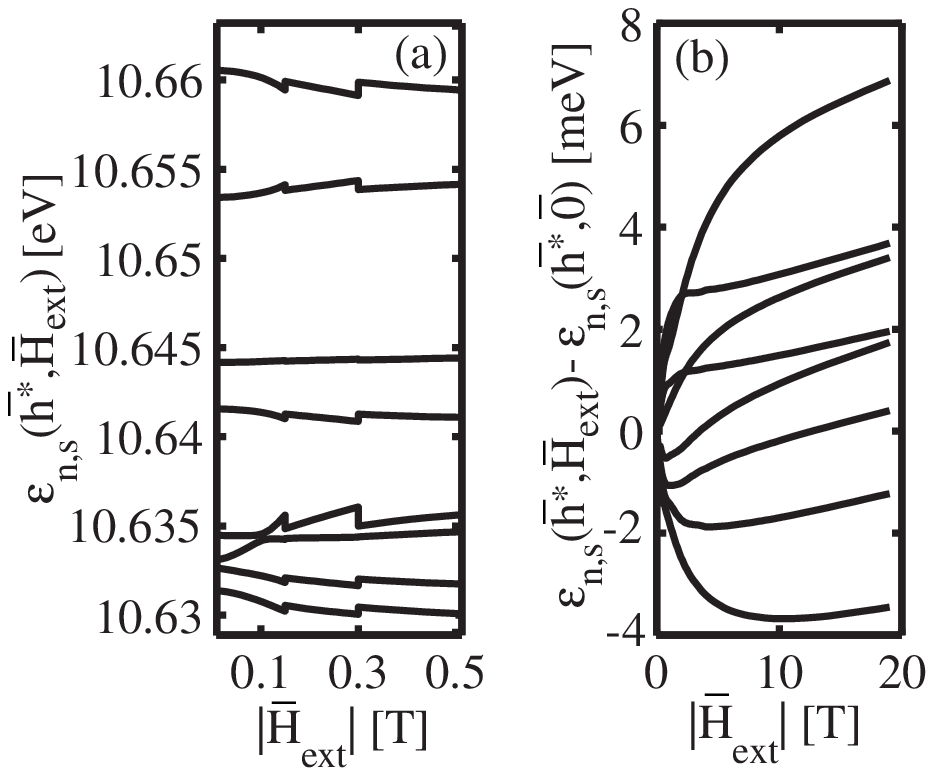}%
\caption{Variation of quasiparticle energy levels
$\epsilon_{n,s}(\vec h^{\star}, \vec H_{\rm ext})$
around the Fermi level
in an external magnetic field.
(a) The levels are calculated assuming that the ground-state 
of the nanoparticle changes in the external field in the way
described in Fig.~[\ref{xyanisdb}]. (b) For magnetic fields
larger than $1$ T, the magnetization starts to develop a non-negligible
component perpendicular to the $xy$ plane, while its projection in the
$xy$-plane is parallel to the component of 
$\vec H_{\rm ext}$ in that plane.}
 \label{mod2}
 \end{figure}
The corresponding field dependence of the quasiparticle is quite complex. 
In the small-field regime ($|\vec H_{\rm ext}| < 1$ T) there is
a hysteretic switching at a certain 
$|\vec H_{\rm ext}| = H_{\rm sw}\approx 1$ T, 
due to an abrupt change of the 
ground-state magnetic moment. There is basically no correlation 
between single particle states before and after reversal.
Notice in particular that the levels can jump
either up or down at $H_{\rm sw}$\footnote{The quasiparticle
energy difference on the opposite sides of the transition is of
the order of the typical quasiparticle anisotropy energy.
For the nanoparticles that we can study numerically this energy
is smaller than the mean level spacing. However for larger particles
the energy difference at the transition can be larger than $\delta$.
In this case the jumps at the switching fields will cause a complete
reshuffling of the quasiparticle levels.}.
Furthermore the quasiparticle energies
have continuous non-monotonic variations, which seem to differ randomly
from level to level. 
In the large field regime
($|\vec H_{\rm ext}| >> H_{\rm sw}$), the quasiparticle energies
depend roughly linearly on $|\vec H_{\rm ext}|$ and their slopes
almost all have the same sign.
In the small-field behavior, the variation of the quasiparticle
energies as a function of the external field 
is due to a combination of the rotation of the particle's
magnetic moment direction $\Omega$,
and Zeeman coupling.  We have shown above (see Fig.~[\ref{eiganis3}])
that the dependence of the quasiparticle energies on $\hat \Omega$
varies randomly from level to level. Thus the complex non-monotonic
behavior at small fields can be understood within our model. At large fields
($|\vec H_{\rm ext}| >> H_{\rm sw}$)
$\hat \Omega$ is slowly twisted toward the field direction and
the Zeeman coupling plays the dominating role in the field dependence.
Moreover, because the Fermi
level lies in a region of predominantly minority-spin energy levels,
it is expected that almost all particle and hole excitations 
around the Fermi level have the same high-field slope\cite{cmc_ahm2000prl}. 

In the discussion above we have assumed, out of simplicity,
that the magnetization is arbitrarily constrained to rotate
in the $zx$-plane.
At weak external fields, however, the magnetization will stay {\it close}
to the $xy$-plane. For a weak external field in 
the $(\theta=\pi/4, \phi=\pi/16)$ direction, the actual dependence of 
the ground-state energy
on the magnetization direction at different field strengths will be the one 
represented in Fig.~[\ref{xyanisdb}]. Consequently, the change of
quasi-particle energies
as a function of the external field will reflect this 
more complicated behavior. We illustrate this point in
Fig.~[\ref{mod2}-a] where, as an example, 
we assume that at zero field
the magnetization points in the $-\hat x+\hat y$ direction.
As the magnetic field increases, there will be a first jump in the
in the quasiparticle-level dependence at $H_{\rm sw}\approx 0.15$ T, when the
magnetization re-orients itself in the $(+\hat x+\hat y)$-direction.
There will a second hysteresis jump in the quasi-particle energies
at $H_{\rm sw}\approx 0.3$ T, when the magnetic moment finally switches
to the $(+\hat x-\hat y)$-direction, corresponding to the 
only metastable configuration left at this field.
For a different starting local minimum at $|\vec H_{\rm ext}|=0$,
the variation of the quasiparticle energies as a function of the external field
will be in general different.
By increasing the magnetic field beyond $1$ T, the magnetization will
start to develop a non negligible component perpendicular to the
$xy$-plane, while its component in the $xy$-plane will become
essentially frozen along the direction of the component of 
$\vec H_{\rm ext}$ in that plane.

The quasiparticle-energy dependence 
on an external magnetic field presented
here, has striking similarities with the 
field dependence of the tunneling resonance energies in single-electron
tunneling experiments\cite{gueron1999,deshmukh2001}. 
Notice that for the hysteretic behavior occurring
as a result of the rotation of the magnetization in the vicinity of
the easy $xy$-plane,
the order of magnitude of the coercivity that we find is close to the
experimental value.
The only discrepancy between the results of our theoretical
model and the experiment is the mean level spacing $\delta_{\rm res}$
of the low-energy excitations.
Since in our model we are only taking into account
Hartree-Fock quasiparticles, the expected level spacing of the
low-lying excitation is approximately $\delta_{\downarrow}$, 
which is a few
meV in our 143-atom nanoparticle and should be of the order of
0.5 meV for a 1500-atom nanoparticle considered in the experiment.

This value is still larger than the value $\delta_{\rm res} \le 0.2$
meV observed experimentally. We believe that a unified and
consistent inclusion of collective spin excitations,
and possibly also the non-equilibrium transport effects proposed
in Refs.~\cite{kleff2001prl,kleff_vdelft2001prb}, 
could resolve this confusion.

\section{Conclusions}
\label{conclusion}

In this article we have investigated the effects of spin-orbit
interactions on the properties of ferromagnetic metal nanoparticles.
In particular, we have focused on their novel microscopic 
magnetocrystalline anisotropy physics, and on hysteresis in the
quasiparticle excitations spectra of metallic nanomagnets.
Our analysis, based on qualitative considerations backed up by 
numerical studies of a generic model for ferromagnetic transition metal 
nanoparticles, provides an understanding of some emergent properties
of their quasi-particle states.  We find two regimes separated by a broad crossover
and characterized by the comparison of several characteristic energy scales. 
For small nanoparticles with fewer than ${\cal N}_a \sim 200$ atoms,
the single particle mean-level spacing $\delta$ is larger 
than spin-orbit induced energy-shifts $\epsilon _{\rm SO}$ in the quasi-particle spectra.
These shifts have the same typical size as the spin-orbit scattering lifetime
broadening energies of very large particles, $\hbar \tau_{\rm SO}^{-1}$.
The quasi-particle levels of small nanoparticles in which 
$\delta \tau_{\rm SO} > \hbar$ evolve in a complicated way as a function
of magnetization orientation and external magnetic field,
with relatively small avoided crossing gaps  and spin-orbit shifts of nearby orbitals
that are nearly uncorrelated.  The size of the avoided crossing gaps is determined by 
matrix element of the spin-orbit coupling operator between quasiparticle energy 
levels that are adjacent in energy.  Surprisingly, even though expectation values
of these matrix elements vanish because of angular momentum quenching, 
typical values between energetically adjacent orbitals are comparable to those 
for orbitals at any position in the spectrum.  Eventually, for nanoparticles
with more than $\sim 1000$ atoms,  the typical avoided
crossing gap estimated in this way becomes comparable to the level spacing
and we expect the crossover to the strong coupling limit is complete. 
Nanoparticles with fewer than ${\cal N}_a \sim 1000$ atoms 
can easily be prepared with current synthesis techniques and systems of experimental 
interest are often in the middle of the crossover between small particle and bulk (weak and strong
spin-orbit coupling) limits.
For example, the nanoparticles investigated by electron tunneling experiments
contain between 50 and 1500 atoms\cite{gueron1999, deshmukh2001}. 

For nanoparticles in the size range we are able to study numerically, 
${\cal N}_{a}$ smaller than a few hundred, we find that the anisotropy energy per atom
displays large changes of order several percent when the electron or atom number
changes by one.  Our analysis allows us to make a connection between
the microscopic model of a metal nanomagnet and more familiar
classical micromagnetic energy functional expressions
which are normally used to interpret the results of
switching-field experiments\cite{jamet2001}. The ground-state energy as a function
of the magnetization direction is characterized by minima separated by
energy barriers. The quasi-particle levels exhibit a complex non-monotonic
behavior and abrupt jumps
when the magnetization direction is reversed by an external magnetic field.
The nanoparticles investigated by electron tunneling experiments
contain between 50 and 1500 atoms\cite{gueron1999, deshmukh2001}. 
The results that we have presented here are therefore
particularly relevant for the interpretation of these experiments.
In particular, we find that the anisotropy fluctuations inferred 
from interpretations of these experiments are indeed to be expected. 
Similarly, the dependence of the tunneling resonances on the external
magnetic field is qualitatively similar to the behavior of the quasi-particle 
excitations of our model, although it appears necessary to 
invoke non-equilibrium
quasiparticle configurations\cite{kleff_vdelft2001prb} 
in the nanoparticle in order to understand the 
size of the quasiparticle level spacing.  

\section{Acknowledgements}
It is a pleasure to thank Mandar Deshmukh and Dan Ralph for several
discussions about their experimental results, 
and Jan von Delft for stimulating and informative interactions. 
C.M.C. would like to acknowledge helpful discussions with L. Samuelson and 
with the participants of the 2001 workshop "Spins in Nanostructures",
at the Aspen Center
for Physics, where part of this work was completed.
A.H.M. acknowledges informative discussions with W. Wernsdorfer and P. Brouwer. 
This work was supported in part by the Swedish Research Council
under Grant No:621-2001-2357 and in part by the National Science Foundation
under Grant DMR 0115947.


\end{document}